\documentclass[aps,prb,twocolumn,groupedaddress,showpacs]{revtex4-1}
\usepackage{amsmath}
\usepackage{amsfonts}
\usepackage{amssymb}
\usepackage{MnSymbol}
\usepackage{xspace}
\usepackage{wasysym}
\usepackage{graphicx}
\usepackage[utf8]{inputenc}
\usepackage[unicode]{hyperref}

\DeclareMathOperator{\Tr}{Tr}

\newcommand{\ie}[0]{i.e.\@\xspace}

\newcommand{\etal}[0]{\textit{et al.}\@\xspace}

\begin{document}

\preprint{}

\title{Kondo screening of spin-charge separated fluxons by a helical liquid}


\author{Manuel Weber}
\email[Electronic address: ]{mweber@physik.uni-wuerzburg.de}
%\thanks{}
\author{Martin Hohenadler}
%\author{Tobias M\"uller}
\author{Fakher F. Assaad}
\affiliation{Institut f\"ur Theoretische Physik und Astrophysik,
Universit\"at W\"urzburg, Am Hubland, 97074 W\"urzburg, Germany}


\date{\today}

\begin{abstract}
The insertion of a magnetic $\pi$~flux into a quantum spin Hall insulator creates four localized, spin-charge separated states:
the charge and spin fluxons with either charge $Q=\pm1$ or spin $S_z=\pm1/2$, respectively.
In the presence of repulsive Coulomb interactions, the charged states are gapped out and a local moment is formed.
We consider the Kane-Mele-Hubbard model on a ribbon with zigzag edges to construct an impurity model
where the spin fluxon is screened by the helical edge liquid.
In the noninteracting model, the hybridization between fluxon and edge states is dominated by the extent of the latter.
It becomes larger with increasing spin-orbit coupling~$\lambda$ but only has nonzero values for even distances between the $\pi$~flux and the edge.
For the interacting system, we use the continuous-time quantum Monte Carlo method,
which we have extended by global susceptibility measurements to reproduce the characteristic Curie law of the spin fluxon.
However, due to the finite extent of the fluxons, the local moment is formed at rather low energies.
The screening of the spin fluxon leads to deviations from the Curie law that follow the universal behavior obtained from a data collapse.
Additionally, the Kondo resonance arises in the local spectral function between the two low-lying Hubbard peaks.
% 
% The insertion of a $\pi$~flux into a quantum spin Hall insulator creates four spin-charge separated states:
% the two charge fluxons with $Q=\pm1$ and the two spin fluxons with $S_z=\pm1/2$.
% In the presence of repulsive Coulomb interactions, the charged states are gapped out and a local moment is formed.
% For both free and interacting systems the fluxons lead to a characteristic Curie law in the magnetic susceptibility.
% We consider the Kane-Mele-Hubbard model on a ribbon with zigzag edges to show that the spinon can be screened by the edge states of a quantum spin Hall insulator.
% At $U=0$ their hybridization is dominated by the extent of the edge states, which becomes larger with increasing spin-orbit coupling~$\lambda$.
% As the fluxons are exponentially localized, it is sufficient to include Hubbard interactions only at lattice sites directly around the $\pi$~flux.
% We have extended the continuous-time quantum Monte Carlo method by global susceptibility measurements that reproduce the Curie law of a free $\pi$~flux even for this reduced interacting system.
% When the spinon is screened by the edge states, we observe deviations from the Curie law for different $U$ and $\lambda$ that follow the universal behavior obtained from a data collapse.
% Moreover, at low temperatures a Kondo resonance arises in the spectral function between two low-lying Hubbard peaks.
\end{abstract}

%\keywords{}

\maketitle

\section{Introduction}

In 1988, Haldane proposed a spinless model for the by then experimentally well established quantum Hall effect that did not rely on an external magnetic field
but instead on a periodic magnetic flux density.\cite{Haldane}
In 2005, Kane and Mele considered the effect of spin-orbit coupling in graphene.
Their model corresponds to a time-reversal invariant generalization of Haldane's model
and exhibits the quantum spin Hall state with a $\mathbb{Z}_2$ topological invariant.\cite{KM1,KM2}
This new state of matter was soon predicted to exist in HgTe quantum wells\cite{Bernevig06}
and was also experimentally realized.\cite{Koenig07}
Since then, the study of topologically nontrivial states has become a very active field, with upcoming subjects like
3D~topological insulators, topological superconductors, or the possibility of observing Majorana fermions in
solid-state systems;
for an introduction to these topics see Ref.~\onlinecite{HasanKane}.

One particular consequence of the nontrivial topology in quantum spin Hall insulators (QSHIs) is the existence of spin-charge
separated states in the presence of a magnetic $\pi$~flux.\cite{Lee07,Weeks07,Ran08,QiZhang08}
Lee~\etal\cite{Lee07} showed that these states are deeply connected with domain walls in one-dimensional soliton models,\cite{JackiwRebbi76,SuSchriefferHeeger79,SuSchriefferHeeger80}
where each soliton leads to one midgap state per spin sector.
Since these states are created by taking half a state from both the valence and the conduction band,
a fully filled valence band lacks a total charge of one---but no spin---in the presence of a soliton.\cite{SuSchriefferHeeger80}
Thus, the four possibilities of occupying the two midgap states lead to four spin-charge separated states
with either nonzero charge or spin quantum number defined within exponential accuracy around the domain wall.\cite{KivelsonSchrieffer82}
A $\pi$~flux inserted into a QSHI
%with $U(1)$ spin symmetry and particle-hole symmetry
also creates one midgap state per spin sector.\cite{Ran08,QiZhang08}
In this context, the exponentially localized\cite{Ran08,QiZhang08} spin-charge separated states are called the charge and spin fluxons.\cite{Ran08}
%and obey nontrivial exchange statistics.\cite{Ran08}

The quantum spin Hall state is robust against weak interactions.\cite{Rachel10,Hohenadler11,Zheng11}
However, when correlation effects are taken into account, identifying the topologically nontrivial state becomes a challenge;
see Ref.~\onlinecite{HohenadlerReview} for a review.
$\pi$~fluxes have been considered as a bulk probe of the $\mathbb{Z}_2$~invariant\cite{Ran08,QiZhang08,Juricic12,Mesaros13,AssaadBercxHohenadler13}
that is also useful in the presence of repulsive electron-electron interactions.
While the charge fluxons are gapped out by interactions, the spin fluxons remain as low-energy states.\cite{Ran08}
Their signatures can be measured using numerical methods like quantum Monte Carlo.\cite{AssaadBercxHohenadler13}
Moreover, static $\pi$~fluxes have been used to construct interacting spin chains,\cite{AssaadBercxHohenadler13} whereas dynamical $\pi$~fluxes have been considered
in interacting topological insulators showing nontrivial exchange statistics.\cite{Ran08,Ruegg12}

In a correlated QSHI, a $\pi$~flux creates a free spin that is exponentially localized.\cite{Ran08,AssaadBercxHohenadler13}
This property suggests interpreting the spin fluxon as a magnetic impurity.
There have been both analytical and numerical studies of a magnetic impurity at the edge of a QSHI
showing the Kondo effect.\cite{Maciejko09,Tanaka11,Maciejko12,Eriksson12,Posske13,Eriksson13,Hu13,Goth13}
The magnetic moment of the impurity gets screened by the electronic bath constituted by the edge states, leading to the formation of a Kondo singlet.\cite{Coleman,Hewson}

In this paper, we study the Kondo screening of a spin fluxon by the helical edge states in the framework of the Kane-Mele-Hubbard model.\cite{Rachel10}
There are some differences to
%ordinary impurity models like
the single-impurity Anderson model\cite{Anderson61}
that complicate the construction of an impurity model.
Usually, a single impurity orbital couples to an electronic bath with a well-defined hybridization parameter,
whereas in our model the impurity is an extended object that is created inside the lattice of the QSHI.
Moreover, we need Hubbard interactions to gap out the charge fluxons and thereby create a free spin.
As long as interactions are not strong enough to destroy the quantum spin Hall phase,\cite{Rachel10,Hohenadler11,Zheng11}
the edge states remain gapless and can, for our purposes, be considered noninteracting.
However, due to the finite extent of the fluxons, we still have to identify a region around the $\pi$~flux
where correlations are necessary to establish the local moment.
%to create the free spin properly.

The organization of this paper reflects the construction of the impurity model.
We first introduce the noninteracting Kane-Mele model in Sec.~\ref{KMPi} and explain how to insert $\pi$~fluxes.
In Sec.~\ref{HybPiE}, we study the hybridization between the fluxon and edge states as a function of distance and spin-orbit coupling.
%We find that it is only nonzero for even distances and increases with increasing spin-orbit coupling.
%These dependencies originate from the edge states, as the hybridization between opposite edges shows the same behavior.
In Sec.~\ref{KondoSpin}, we add repulsive Hubbard interactions around the $\pi$~flux.
Using the continuous-time quantum Monte Carlo method in the weak-coupling interaction expansion (CT-INT),\cite{Rubtsov05}
we first demonstrate the formation of a local moment and then the Kondo screening of the spin fluxon.
Evidence for the Kondo effect is provided by showing a data collapse of the static spin susceptibility
and the appearance of the Kondo resonance in the local spectral function.
Section~\ref{conclusions} contains our conclusions.
Finally, the Appendix contains an introduction to the CT-INT method with our extension to global susceptibility measurements.

\section{Kane-Mele model with \texorpdfstring{$\pi$}{Pi}~fluxes\label{KMPi}}

To study $\pi$~fluxes in a noninteracting QSHI, we consider the Hamiltonian of the Kane-Mele model\cite{KM1,KM2} at half filling,
\begin{align}
\hat{H}_{\mathrm{KM}} = 
-t \sum_{\langle i,j \rangle} \tau_{i,j} \hat{c}^{\dagger}_{i} \hat{c}^{\phantom{}}_{j}
+ \mathrm{i} \lambda \sum_{\llangle i,j \rrangle} \tau_{i,j} \nu_{i,j} \hat{c}^{\dagger}_{i} \sigma_z \hat{c}^{\phantom{}}_{j}.
\label{KM_Ham}
\end{align}
We use the spinor notation $\hat{c}_{i} = ( \hat{c}_{i,\uparrow}, \hat{c}_{i,\downarrow}  )^{\top}$
with the operator $\hat{c}_{i,\sigma}$ annihilating an electron in a Wannier state at lattice site~$i$ with spin~$\sigma$.
The model is defined on the honeycomb lattice, with summation indices running over nearest and next-nearest neighbors
denoted by $\langle\bullet\rangle$ and $\llangle\bullet\rrangle$, respectively.
The factor $\tau_{i,j} = \pm 1$ encodes additional signs resulting from the $\pi$~fluxes.
The original Kane-Mele model is recovered by setting $\tau_{i,j} = 1$.

In addition to the tight-binding Hamiltonian of graphene, Eq.~(\ref{KM_Ham}) contains a complex next-nearest-neighbor hopping term due to spin-orbit coupling.\cite{KM1,KM2}
Its sign depends both on the spin and the direction of the hopping, as encoded in the Pauli spin matrix~$\sigma_z$ and the factor $\nu_{i,j}=\pm1$.
We choose $\nu_{i,j}=+1$ when the hopping from site~$j$ to site~$i$ makes a turn to the left and $\nu_{i,j}=-1$ otherwise.
Spin-orbit coupling reduces the $SU(2)$ spin symmetry to a $U(1)$ symmetry.
%which remains unbroken as we do not consider a Rashba coupling.
In this form, the Hamiltonian~(\ref{KM_Ham}) is equivalent to two copies of the Haldane model\cite{Haldane} that both obey particle-hole symmetry.

\begin{figure}[htbc]
 \centering
\includegraphics[width=\linewidth]{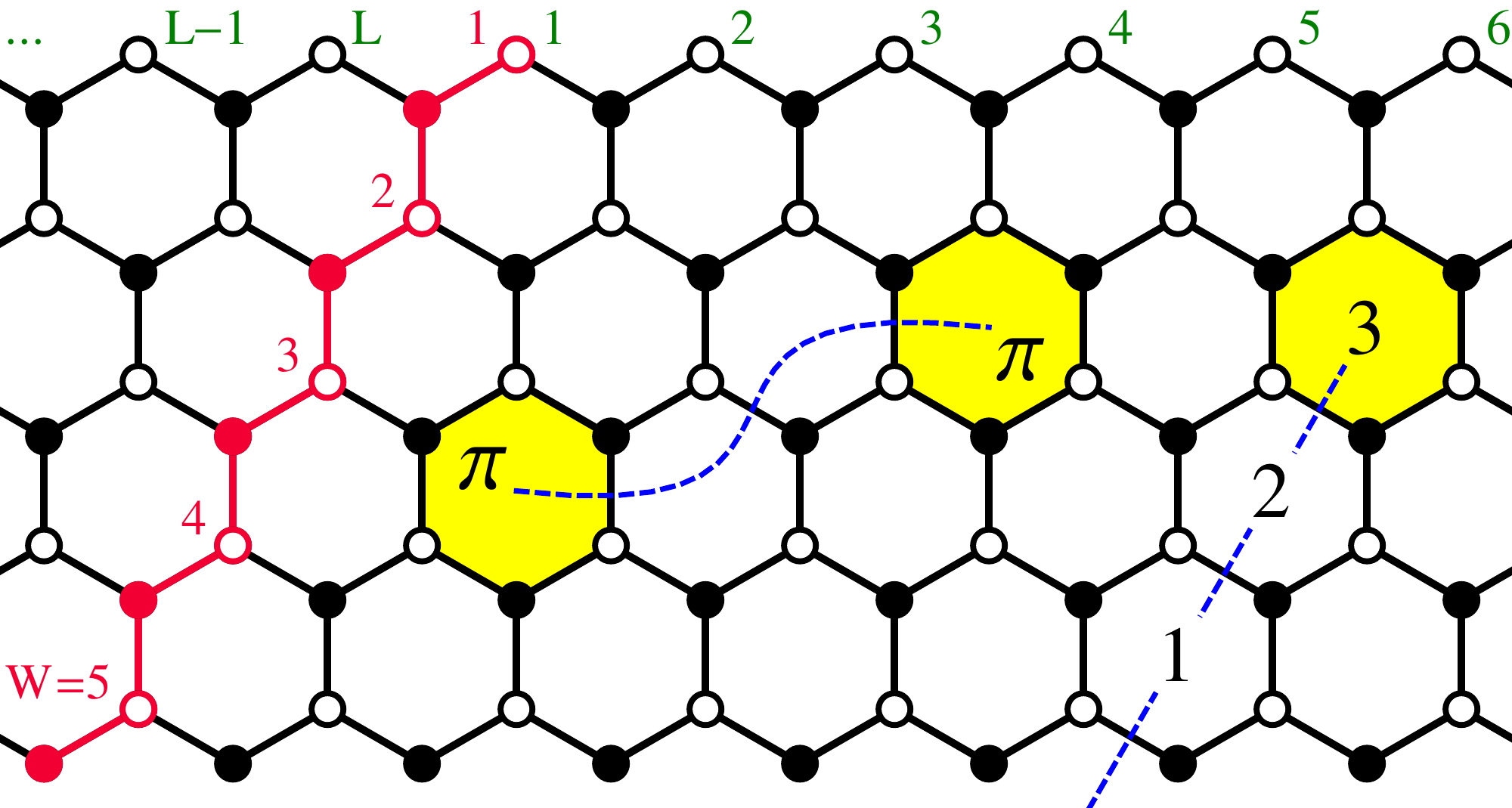}
 \caption{(Color online) A honeycomb ribbon with zigzag edges, length~$L$, and width~$W$.
                         %$W$ counts the number of points in the unit cell (red) that belong to one of the two sublattices.
                         $\pi$~fluxes (yellow) can either be included in pairs connected by a string (blue) or individually
                         with a string cutting one of the edges.
                         The distance~$d$ between the $\pi$~flux and the edge is given by a natural number, as shown in the figure.
              }
 \label{ribbonpiflux}
 \end{figure}

On the honeycomb lattice, a $\pi$~flux corresponds to a magnetic flux of strength $\pi$ threading a single plaquette~$\hexagon$.
The phase acquired by an electron moving around such a $\pi$~flux can be encoded in a string originating in the associated plaquette
and giving a factor of $\tau_{i,j}=-1$ for each hopping element in $\hat{H}_{\mathrm{KM}}$ that crosses the string.
Given periodic boundary conditions, $\pi$~fluxes can only be inserted in pairs at the end points of a string,
whereas open boundary conditions allow us to insert a single $\pi$~flux,
%since the second end of the string can be placed out of the system.
as shown in Fig.~\ref{ribbonpiflux}.
The string itself can be chosen arbitrarily, as different configurations are related by gauge transformations.
Since the $\pi$~flux is strictly localized at its plaquette, it does not break time-reversal or particle-hole symmetry.

In the following, we will mainly consider a ribbon geometry with zigzag edges, as shown in Fig.~\ref{ribbonpiflux}.
Its unit cell contains $2W$ lattice sites, where $W$ defines the width of the ribbon.
The length of the ribbon is given by the number~$L$ of unit cells in the periodic direction.
The distance between two unit cells is denoted by $a=\sqrt{3}a_0$, where $a_0$ is the lattice constant. 
As the ribbon has open boundary conditions in one direction, we can insert a single $\pi$~flux at distance~$d$ to the edge,
where $d$ is defined as in Fig.~\ref{ribbonpiflux}.
%In the following, we will consider a single $\pi$~flux at distance~$d$ to the edge of a ribbon with zigzag edges, as given in Fig.~\ref{ribbonpiflux}.
%The size of the ribbon is given by its width~$W$ and length~$L$,
%where $2W$ is the number of lattice sites per unit cell and $L$ the number of unit cells in the periodic direction.

\section{Determining the hybridization between fluxon and edge states\label{HybPiE}}

Since the Kondo impurity considered here is created by a $\pi$~flux through the honeycomb lattice,
we do not have a simple hybridization parameter that controls the coupling between impurity and bath.
However, the hybridization can only depend on the extent of both the fluxon states and the edge states,
and their distance.
%Thus, we will first study how the latter depends on the spin-orbit coupling~$\lambda$ and the distance.

\subsection{Coupling between two zigzag edges\label{coupling_edges}}

On an infinite system, the helical edge states of a QSHI show a crossing in the energy spectrum at the time-reversal invariant point $k = \pi/a$.
However, on a finite ribbon, a gap $\Delta\epsilon \propto \exp(-W/\xi_E)$ opens at $k = \pi/a$ that vanishes exponentially with the width $W$ of the ribbon.\cite{Zhou08}
It is generated by the finite overlap of the edge states at the two opposite edges.
Vice versa, the decay length $\xi_E$ will give a good estimate of the extent of the edge states.

\begin{figure}[tbc]
 \begin{tabular}{cc}
 \hspace{-0pt}\includegraphics[trim=0.55cm 0.3cm 0cm 0cm, clip=true, height=0.51\linewidth]{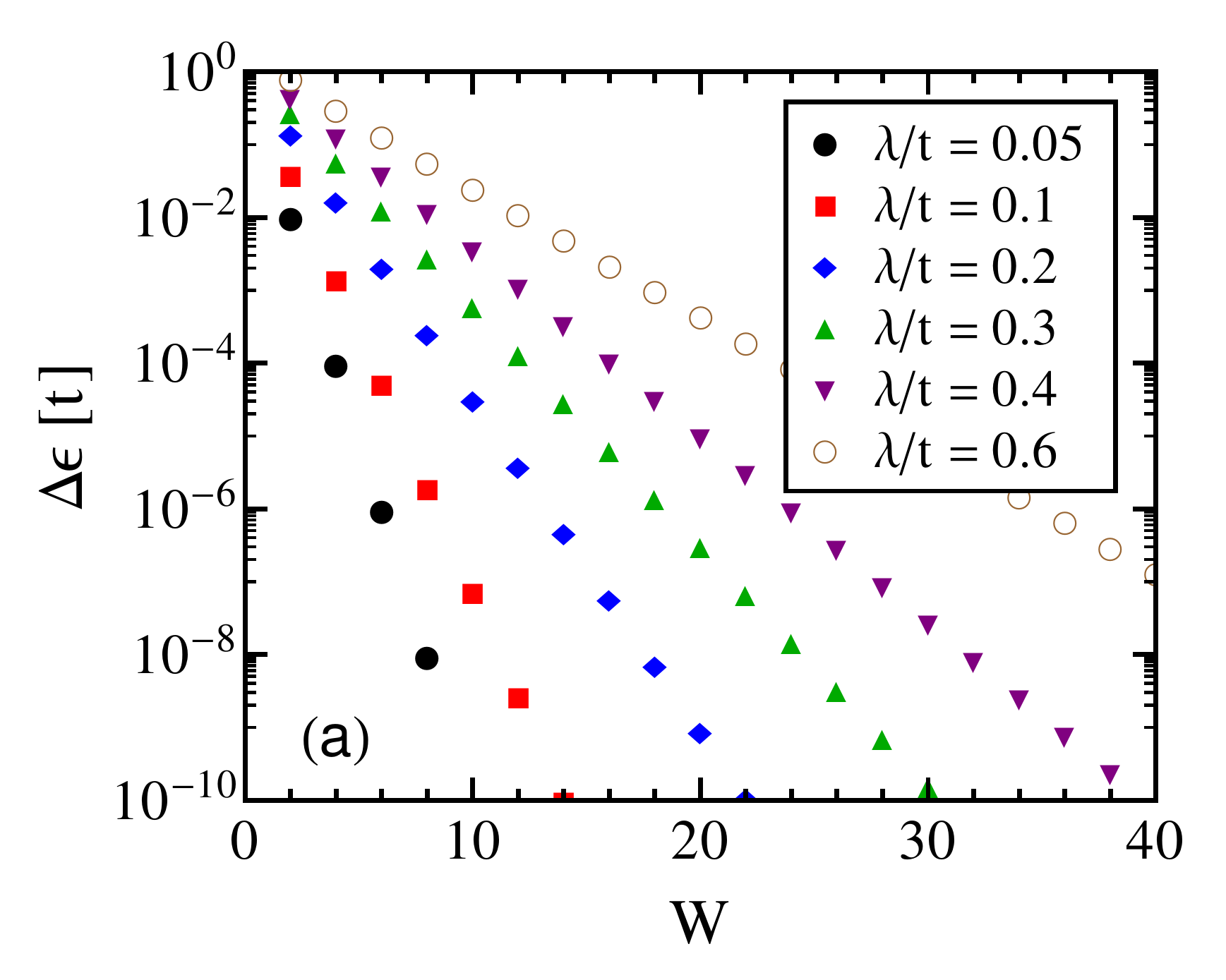} & \hspace{-0.3cm}\includegraphics[trim=4cm 0.1cm 4cm 0.5cm, clip=true, height=0.49\linewidth]{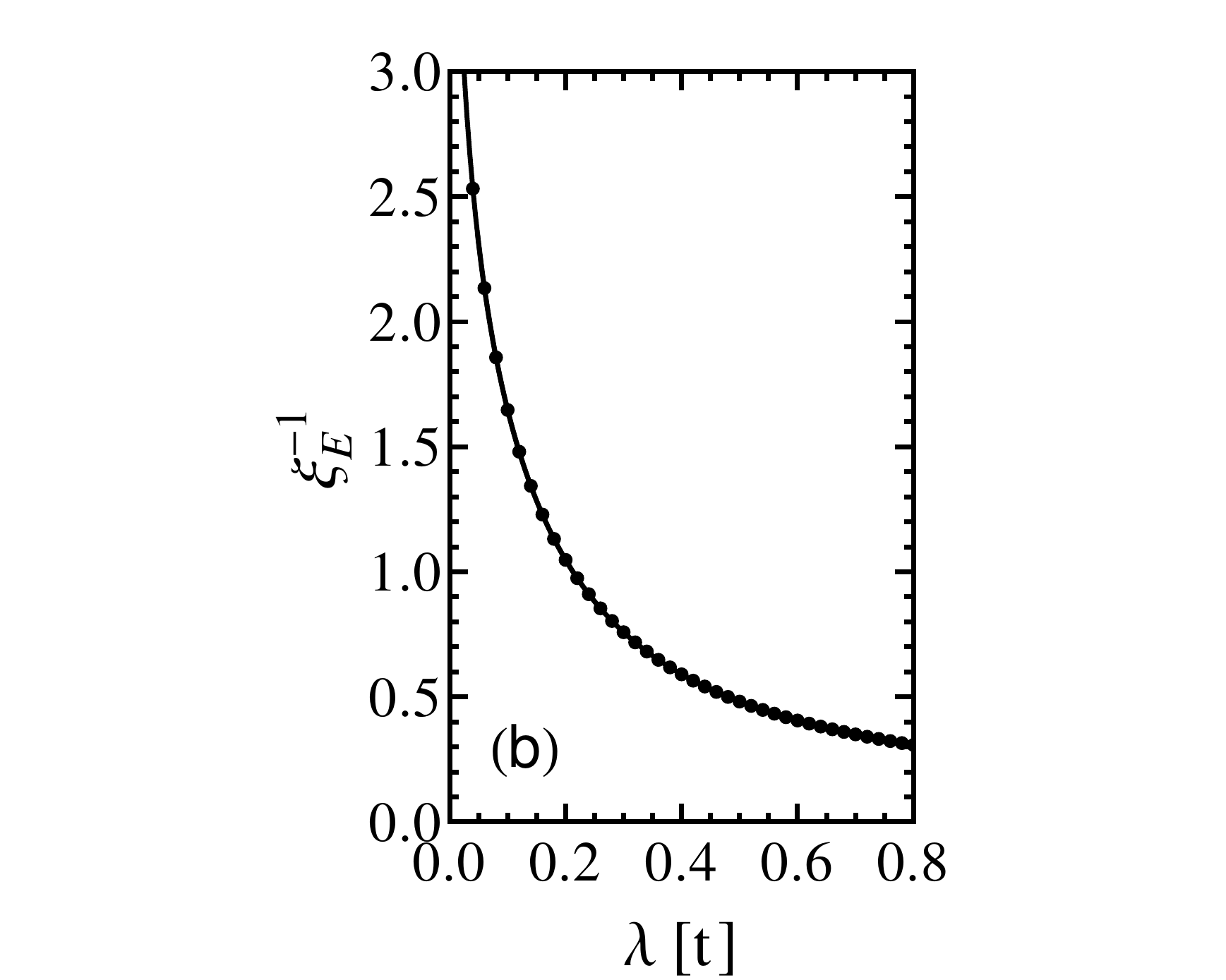}
 \end{tabular}
\caption{(Color online) (a)~The energy gap $\Delta\epsilon$ at $k=\pi/a$ for different $\lambda$ as a function of the width~$W$ of the ribbon.
         (b)~The inverse decay length~$\xi_E^{-1}$ as a function of~$\lambda$ together with the exact result\cite{vandenBrink13} (solid line).}
\label{edgedecay}
\end{figure}

%The finite distance between the zigzag edges of the Kane-Mele model also leads to the opening of an energy gap~$\Delta\epsilon$,
%as shown in Fig.~\ref{edgedecay}a for different values of the spin-orbit coupling~$\lambda$.
Figure~\ref{edgedecay}(a) shows $\Delta\epsilon$ for the Kane-Mele model on a finite ribbon with zigzag edges
for different values of the spin-orbit coupling~$\lambda$.
%but only for even values of the width~$W$.
From the exponential decay one can estimate the inverse decay length~$\xi_E^{-1}$ which monotonically decreases as a function of $\lambda$, see Fig.~\ref{edgedecay}(b).
It is given exactly by $\xi_E^{-1} = \operatorname{Re} \cosh^{-1} ( \mathrm{i}t/4\lambda )$.\cite{vandenBrink13}
According to Fig.~\ref{edgedecay}(b), the extent of the edge states becomes larger with increasing $\lambda$.
However, Fig.~\ref{edgedecay}(a) shows that the dispersion remains gapless for odd $W$, as already observed in Refs.~\onlinecite{Zarea09,Chung14}.
Thus, there is only a coupling between opposite edge states for even $W$.

\subsection{Coupling between two fluxons\label{coupling_fluxons}}

A single $\pi$~flux deep in the bulk of a QSHI leads to one energy mode per spin sector lying exactly at zero energy as particle-hole symmetry is conserved.\cite{Lee07,Ran08}
However, on a torus geometry, one must include two $\pi$~fluxes, and the exponentially localized fluxon states will always have a finite spatial overlap.
Their hybridization causes a splitting of the zero modes by $\pm\Delta\epsilon$, exactly as in the case of solitons in the Su-Schrieffer-Heeger model.\cite{JackiwSchrieffer81}

\begin{figure}[htbc]
 \begin{tabular}{cc}
 \hspace{-0pt}\includegraphics[trim=0.3cm 0cm 0cm 0cm, clip=true, height=0.54\linewidth]{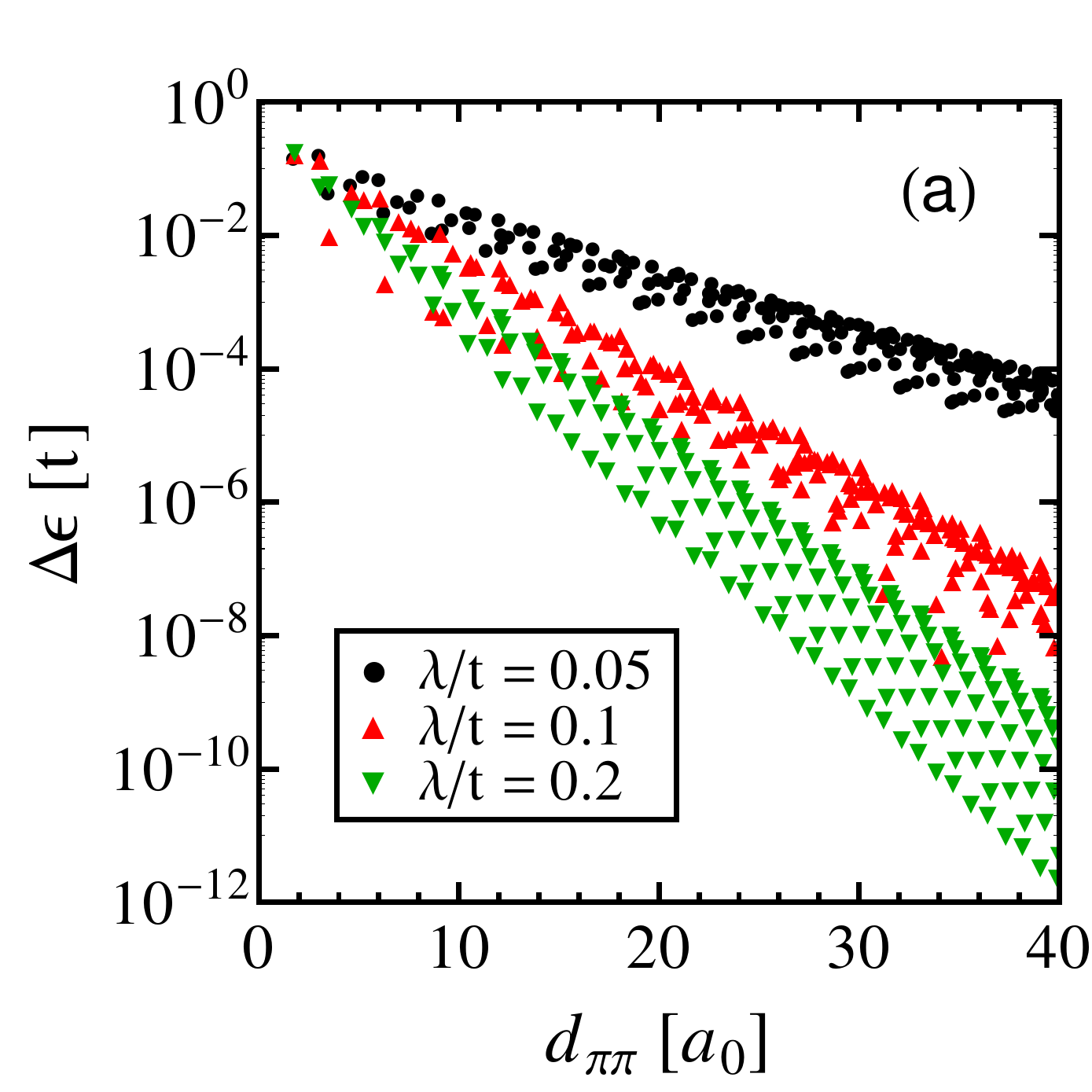} &
 \hspace{-0pt}\includegraphics[trim=3.0cm 0cm 0cm 0cm, clip=true, height=0.54\linewidth]{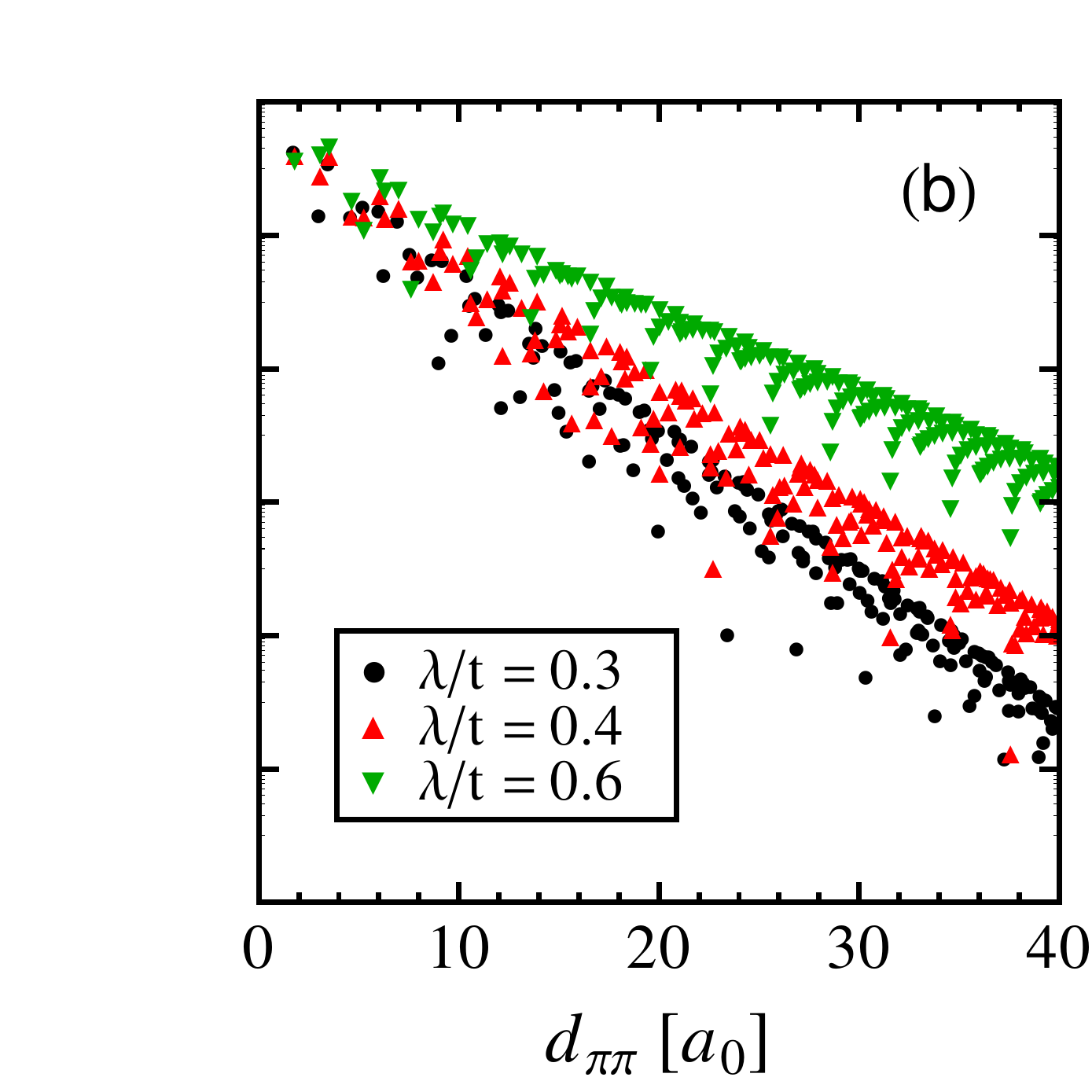}
 \end{tabular}
 \caption{(Color online) The energy splitting~$\Delta\epsilon$ due to the finite distance~$d_{\pi\pi}$ between two $\pi$~fluxes
          on a $50\times50$~lattice is shown for weak~(a) and strong~(b) spin-orbit coupling~$\lambda$.
          %The insets depict $\Delta\epsilon$ for small distances.
          }
\label{deltaepsilon}
 \end{figure}

As in Sec.~\ref{coupling_edges}, we can use $\Delta\epsilon$ to study how the extent of the fluxon states depends on the spin-orbit coupling~$\lambda$.
To this end, we fix the position of one $\pi$~flux and determine $\Delta\epsilon$ for all the possible positions of the second $\pi$~flux.
The results are shown in Fig.~\ref{deltaepsilon} as a function of the distance~$d_{\pi\pi}$---measured between the centers of the flux-threaded hexagons---and for different
values of $\lambda$.
Keeping $\lambda$ fixed, one observes strong variations of $\Delta\epsilon$ for flux separations
with similar $d_{\pi\pi}$ but different directions on the lattice.
%for almost identical distances,
%as the two $\pi$~fluxes are oriented along different directions in the lattice.
The sets of points for different $\lambda$ are clearly bounded from above and decay exponentially for large distances.
The inverse decay length~$\xi_{\pi}^{-1}$ has its maximum around $\lambda/t = 0.2$, where the fluxons are maximally localized.
However, for very small distances there is no systematic dependence on $\lambda$.
%The same information can be obtained by looking directly at the fluxons' wave functions.

\subsection{Hybridization between fluxon and edge states}

Having analyzed how the extent of both edge states and fluxons depends on the spin-orbit coupling~$\lambda$,
we can insert a $\pi$~flux near the edge of the ribbon.
To determine the hybridization between fluxon and edge states as a function of distance~$d$ and spin-orbit coupling~$\lambda$,
we look for a signature in the static spin susceptibility
\begin{align}
\chi_S = \beta \left( \langle \hat{S}_z^2 \rangle - \langle \hat{S}_z \rangle^2 \right),
\label{spinsus}
\end{align}
with $\hat{S}_z = \sum_{i} \hat{c}_i^{\dagger} \sigma_z \hat{c}_i$.
For an independent $\pi$~flux, we expect a Curie law $\chi_S = 1/2k_B T$ at low temperatures,\cite{AssaadBercxHohenadler13}
whereas for the edge states the susceptibility approaches a constant contribution
$\chi_S = g(\epsilon_F)L$ determined by the normalized density of states $g(\epsilon_F) = 2a/\pi v_F$ at the Fermi level~$\epsilon_F$.
To probe the hybridization between fluxon and edge states, we look for a deviation from the Curie law.
Thus, it is useful to define $\Delta\chi_S$ as the difference between the susceptibility of a ribbon with and without a $\pi$~flux.
%For the noninteracting Hamiltonian $\hat{H}_{\mathrm{KM}}$ the charge susceptibility~$\chi_C$, where $\hat{S}_z$ has to be exchanged with the total particle number~$\hat{N}$,
%always gives the same result, thus we abbreviate the susceptibility with $\chi$.

\subsubsection{Dependence on the distance \texorpdfstring{$d$}{d}}

\begin{figure}[htbc]
 \hspace{-0pt}\includegraphics[trim=0cm 0cm 0.5cm 0cm, clip=true, width=\linewidth]{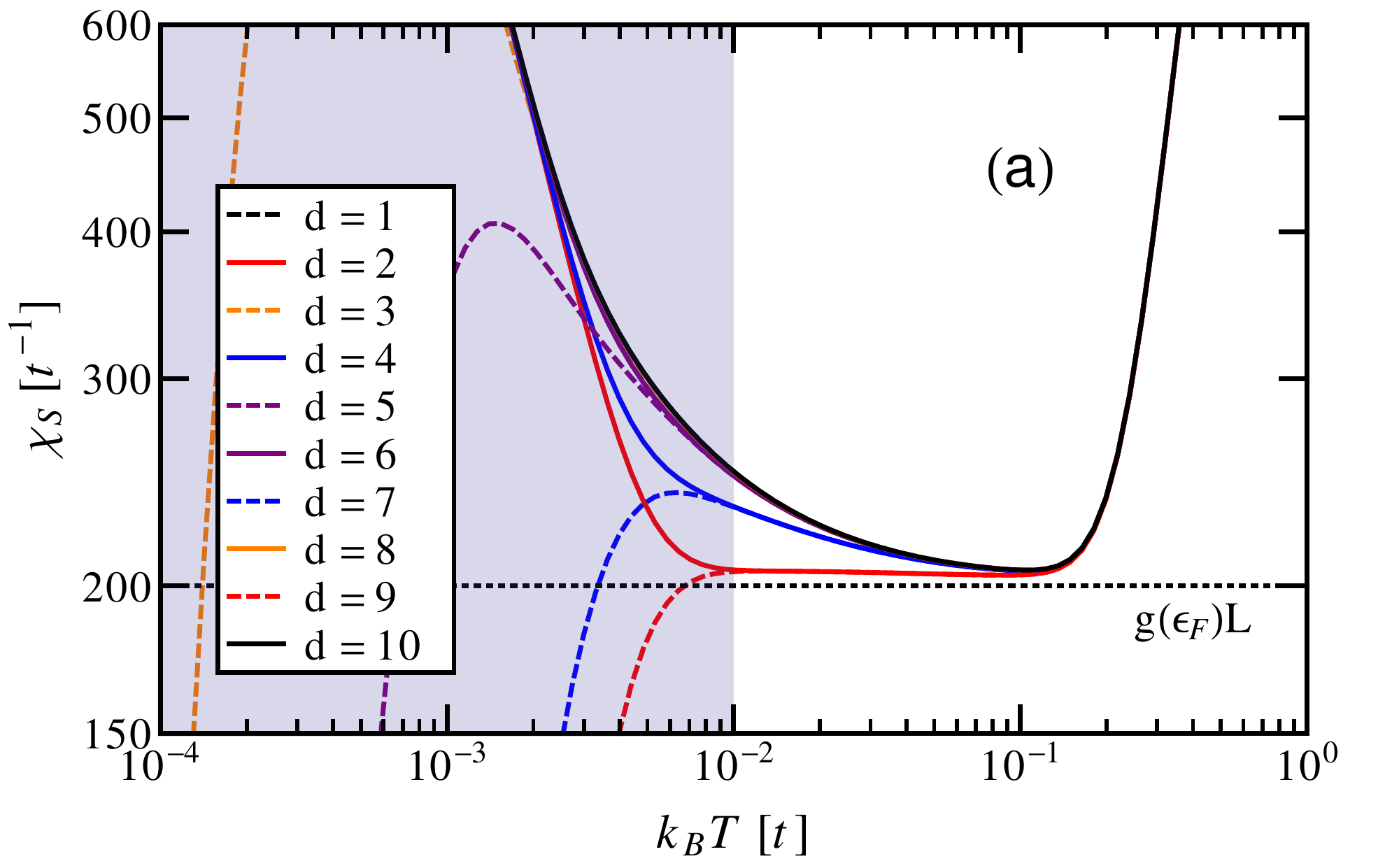} \\
 \hspace{-0pt}\includegraphics[width=0.96\linewidth]{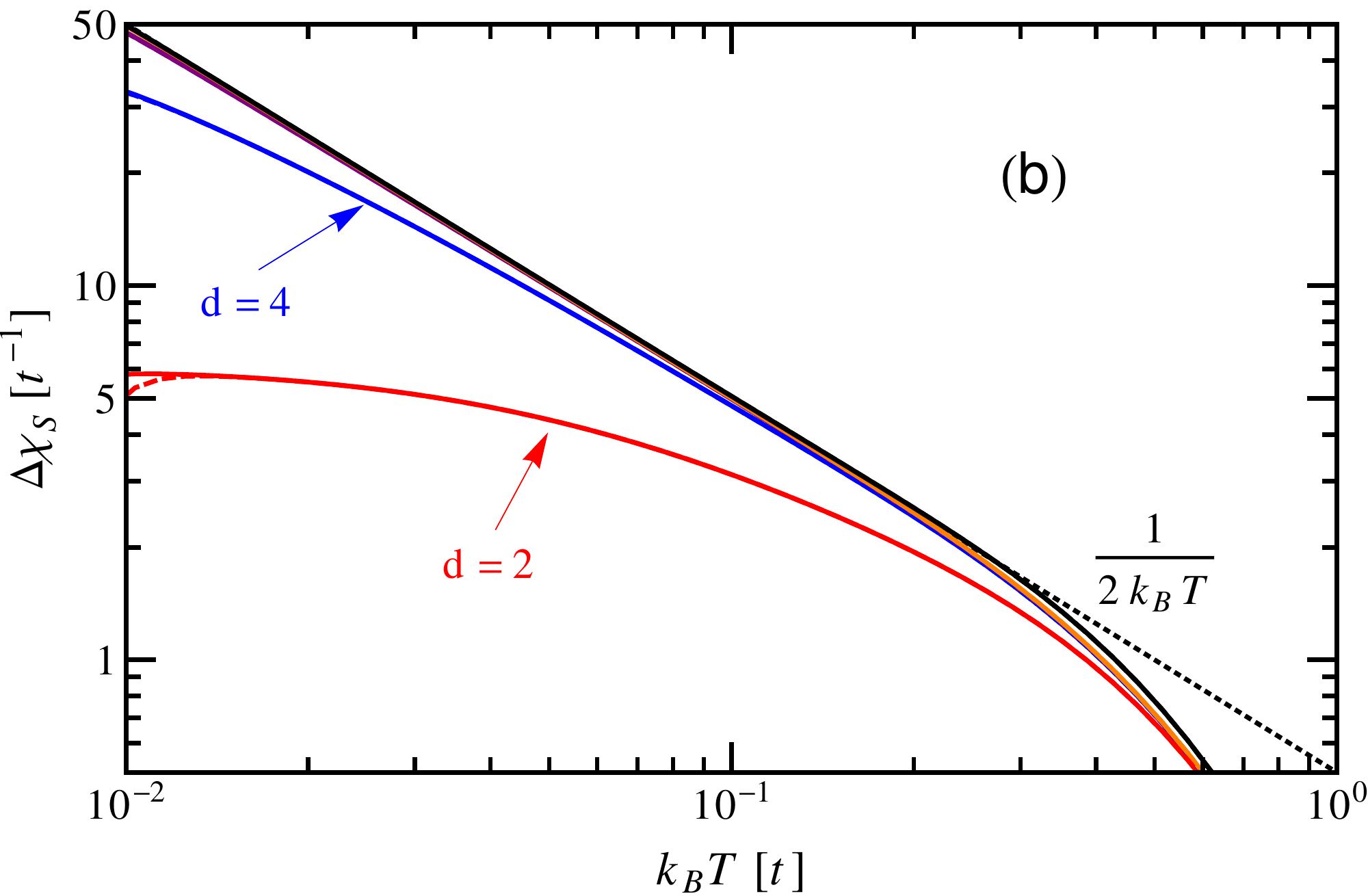}
\caption{(Color online) (a)~The spin susceptibility~$\chi_S$ of a ribbon with~$W=11$, $L=400$ and~$\lambda/t=0.4$
             is shown for all possible distances~$d$ of a $\pi$~flux from the edge.
             For physically equivalent distances $d$ and $W-d$, $\chi_S$ is marked by the same color
             and only differs within the shaded region of finite-size effects, because our gauge choice breaks inversion symmetry around the center of the ribbon.
             For even or odd $d$, $\chi_S$ is drawn as a solid or dashed line, respectively.
             Note that for odd $W$ an even distance to one of the edges corresponds to an odd distance to the other edge.
             %For distances $d$ and $W-d$, $\chi_S$ is marked by a solid or a dashed line of the same color, as the distance to the nearest edge is the same.
             %The inset shows the temperature range where only the contributions from the edge states and the $\pi$~flux are relevant.
             The expected susceptibility of the edges, $g(\epsilon_F)L$, is drawn as a dotted line.
         (b)~To see the Curie law~$\chi_S=1/2k_BT$, we plot the difference $\Delta\chi_S$ of the susceptibilities with and without a $\pi$~flux.
             %At odd distances there is no deviation, but at even distances it becomes larger with decreasing~$d$.
             %Note that there is no difference between coupling to the left or right edge inside the thermodynamically well-defined temperature range.
}
\label{suszribpi}
\end{figure}

Figure~\ref{suszribpi}(a) shows the spin susceptibility~$\chi_S$ of a ribbon of width~$W=11$ and for~$\lambda/t=0.4$ with a $\pi$~flux at different distances~$d$ from the edge.
As the temperature is lowered, the bulk contribution vanishes and $\chi_S$ starts to slowly increase.
There is a dependence on the distance~$d$ which must be a consequence of the hybridization between the fluxon and edge states.
To get rid of the constant contribution of the edge states, we consider $\Delta\chi_S$ in Fig.~\ref{suszribpi}(b).
For odd distances~$d$, the susceptibility approaches the Curie law of a free $\pi$~flux, indicating that there is no hybridization.
However, for even $d$, the fluxon and edge states hybridize and $\Delta\chi_S$ shows a clear deviation from the Curie law for $d=2$ and $d=4$.
As expected, the hybridization becomes smaller with increasing distance to the edge.
Note also that $d=5$ deviates a little from the Curie law, as it corresponds to a distance $d=6$ to the other edge.

When performing susceptibility measurements, one has to be aware of finite-size effects.
In particular, energy scales that are lower than the resolution of the energy eigenvalues cannot be resolved.
In our case, this threshold appears at $k_B T / t \approx 0.01$, as demonstrated by Fig.~\ref{suszribpi}(a).
However, similar to Secs.~\ref{coupling_edges} and \ref{coupling_fluxons}, an analysis of the zero-energy eigenvalues will give some more information
about the hybridization between fluxon and edge states.
Consider a ribbon of odd width~$W$ and even length~$L$, but without a $\pi$~flux.
Since there is a $k$~point at $k=\pi/a$ and no coupling between the edges, one zero-energy eigenvalue per spin sector appears for each edge.
The inclusion of a single $\pi$~flux makes it necessary to lay the string out of the system, leading to antiperiodic boundary conditions for one of the edges.
Thus, the $k$~points of this edge are shifted by $\Delta k = \pi/La$, and the corresponding zero-energy eigenvalue is gone.
Nevertheless, we still find two zero-energy eigenvalues in the spectrum, one for the second edge and one for the $\pi$~flux, as long as these two do not couple.
This is the case for odd distances of the $\pi$~flux to the second edge
and results in a Curie-like finite-size law in Fig.~\ref{suszribpi}(a).
Even distances to this edge will instead again lead to a finite $\Delta\epsilon$, such that on a finite lattice $\chi_S\rightarrow0$ for $T\rightarrow0$.

The analysis of the zero-energy eigenvalues emphasizes that at odd distances the hybridization is not only much smaller than for even $d$, but it is exactly zero.
This odd-even effect can be related to the one we have observed for the coupling between two edge states as a function of the width~$W$.
We adopt the soliton interpretation of the fluxon states given by Lee \etal\cite{Lee07}
Consider a semi-infinite ribbon, which we cut parallel to the edge.
Along this cut, gapless edge states appear that only couple to the original edge states if the cut creates a ribbon of even width~$W$.
For odd $W$, reconnection of the bonds with weaker coupling strength opens a mass gap~$m$ in the counterpropagating edge states without affecting the original edge state.
Reconnecting the bonds with a sign change creates a mass gap that interpolates between $\pm m$.
According to the Jackiw-Rebbi model,\cite{JackiwRebbi76} a solitonic midgap state appears in the energy spectrum which corresponds to the fluxon created by a $\pi$~flux.
As there was no coupling between the edges, the fluxon created as a soliton in a quasi-one-dimensional system cannot couple to the edge at odd distance~$d$ either.

%Reconnecting the cut bonds with weaker coupling strength will open a mass gap~$m$  and inverting the coupling

\subsubsection{Dependence on the spin-orbit coupling \texorpdfstring{$\lambda$}{lambda}}

\begin{figure}[b]
 \begin{tabular}{cc}
 \hspace{-0pt}\includegraphics[trim=0.2cm 0cm 0.1cm 0cm, clip=true, height=0.53\linewidth]{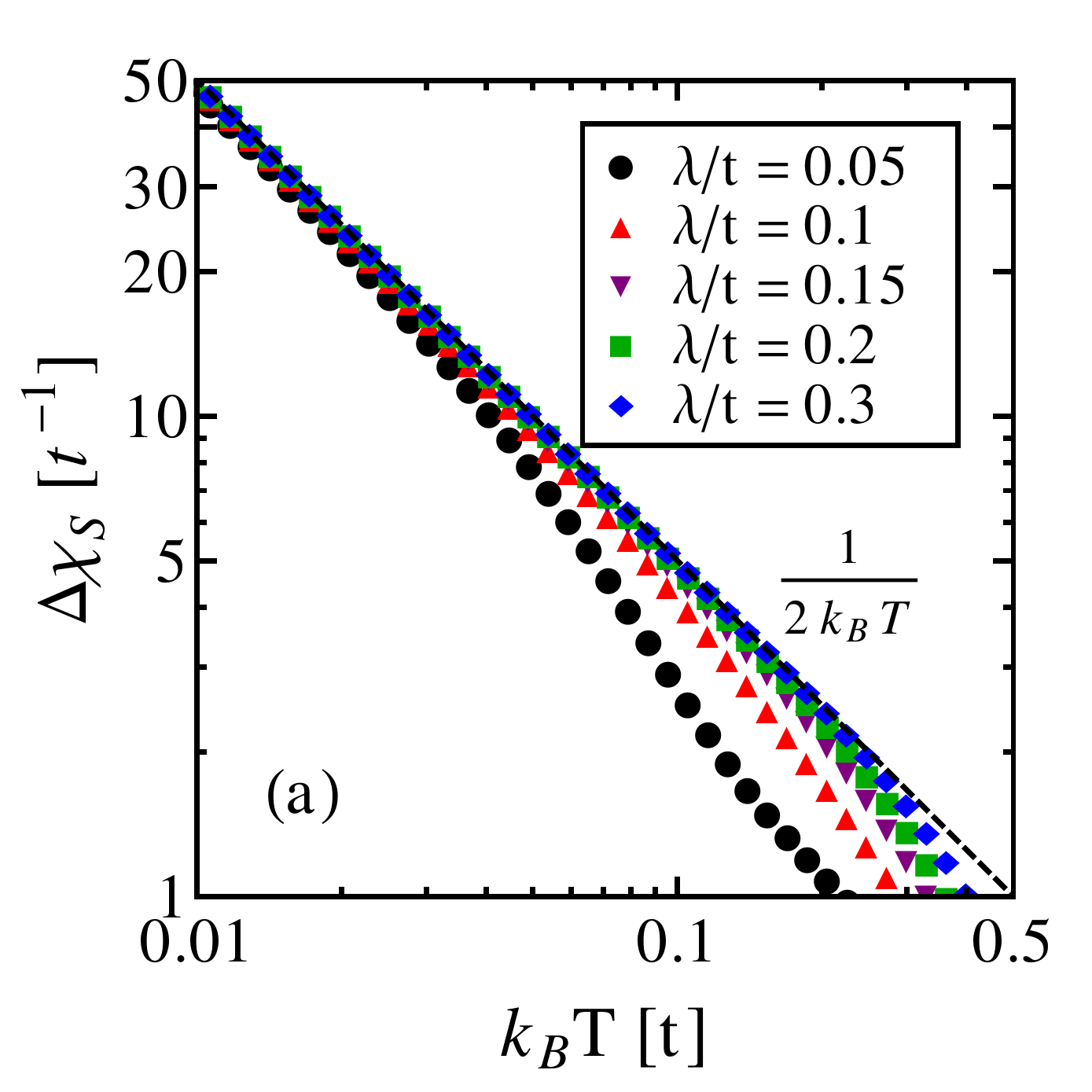} &
 \hspace{-5pt}\includegraphics[trim=0.2cm 0cm 0.1cm 0cm, clip=true, height=0.53\linewidth]{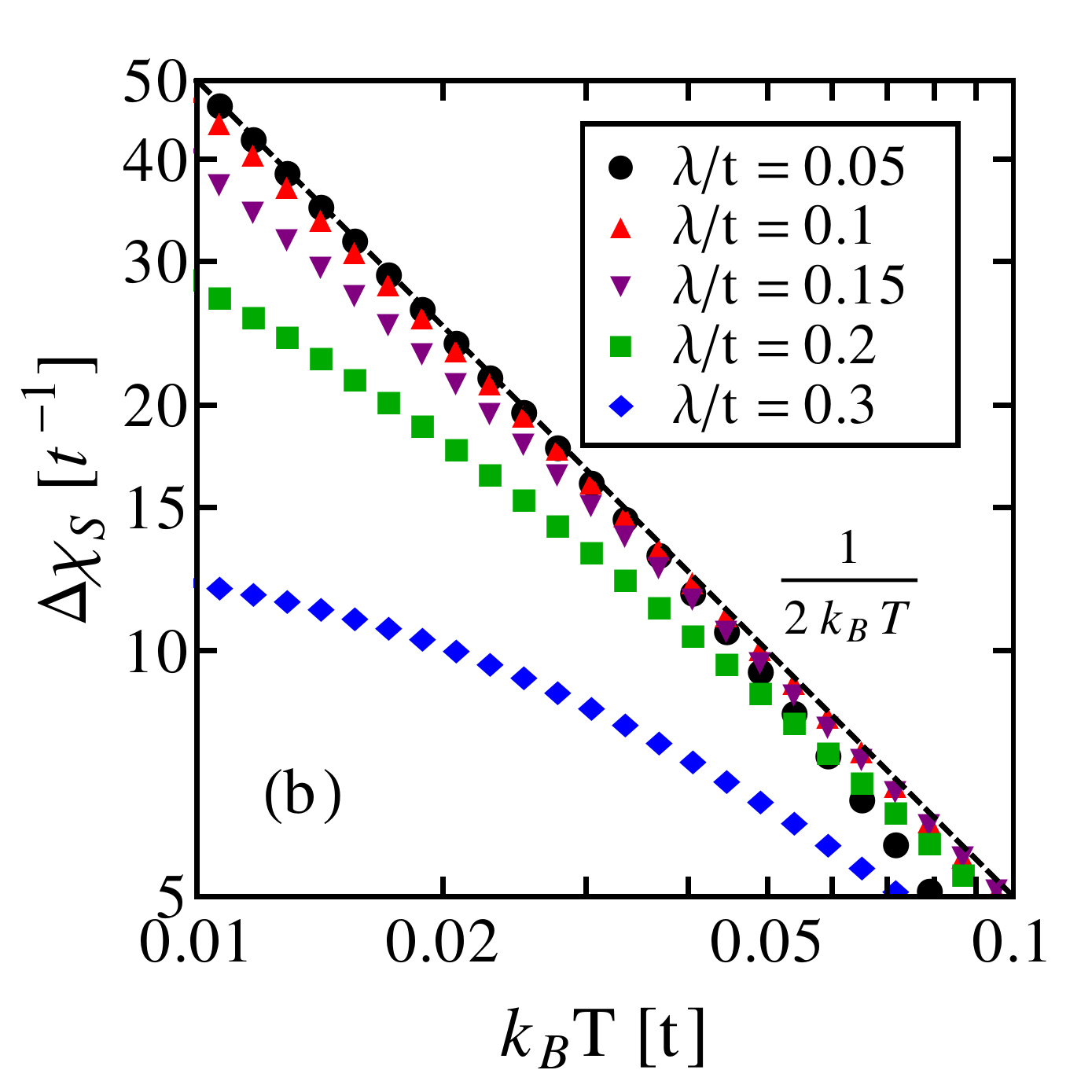}
 \end{tabular}
\caption{(Color online) Spin susceptibility~$\Delta\chi_S$ for different $\lambda$ at distances $d=1$~(a) and $d=2$~(b) of a $\pi$~flux from the edge.
                        The dashed line shows the Curie law~$1/2k_BT$.
                        Here, $W=10$ and~$L=500$.}
\label{abstand}
\end{figure}

To maximize the hybridization between fluxon and edge states, we vary the spin-orbit coupling~$\lambda$ for $d=1$ and $d=2$.
The results are shown in Fig.~\ref{abstand}.
%The observation of the Kondo screening requires to choose appropriate parameters that maximize the hybridization.
%Therefore, we take a closer look
At $d=1$, the susceptibility $\Delta\chi_S$ approaches the Curie law $1/2k_BT$ for each~$\lambda$.
For decreasing $\lambda$, convergence of $\Delta\chi_S$ to the Curie law occurs at lower temperatures,
as the bulk band gap becomes smaller.
We do not observe any hybridization at $d=1$ for any value of~$\lambda$.
In contrast, for $d=2$, there is a hybridization between fluxon and edge states.
While for $\lambda/t=0.05$ the hybridization cannot be observed on the temperature scale shown in Fig.~\ref{abstand}(b),
it becomes larger with increasing $\lambda$, and $\Delta\chi_S$ shows a stronger deviation from the Curie law.
According to the odd-even effect in the distance~$d$, this behavior will generalize to larger distances.

To get a qualitative understanding of the hybridization at even distances~$d$
as a function of $\lambda$,
we have studied the extent of both fluxons and edge states on their own.
We found that the decay length~$\xi_{\pi}$ for a fluxon has a minimum around $\lambda/t=0.2$,
whereas for the edge states $\xi_{E}$ increases monotonically with increasing~$\lambda$.
Since the hybridization between fluxon and edge states also increases monotonically with $\lambda$,
we assume that it is dominated by the extent of the edge states.

\section{Kondo screening of the spin fluxon\label{KondoSpin}}

In the presence of repulsive electron-electron interactions, the fourfold degeneracy of the fluxon states is lifted
because the charge degrees of freedom are gapped out.\cite{Ran08}
A Kramers pair of spin fluxon states remains at low energies and constitutes a free spin.
%In the vicinity of the edge states of a QSHI this free spin will be screened by the helical liquid.
If the distance of the $\pi$~flux from the edge is sufficiently small, the free spin will be screened by the helical liquid.

To study the Kondo screening of the spin fluxon, we consider the particle-hole symmetric Hubbard interaction
\begin{align}
 \hat{H}_U = \frac{U}{2} \sum_{i\in\hexagon} \left( \hat{c}^{\dagger}_{i} \hat{c}^{\phantom{}}_i -1 \right)^2.
\label{Hint}
\end{align}
%that is only introduced at the six lattice sites of the flux-threaded plaquette.
Since the edge states do not show substantial changes up to intermediate interaction strengths,\cite{Hohenadler11}
we will include the Hubbard interactions only at the six sites threaded by the $\pi$~flux to create the free spin.
%In Sec.~\ref{localmoment} we will show that this is sufficient.
As shown below, this setup is sufficient for our purposes and significantly reduces the numerical effort.

We use the CT-INT method introduced by Rubtsov~\etal\cite{Rubtsov05} to treat $\hat{H}_U$ numerically exactly in the weak-coupling interaction expansion.
A short introduction to this method can be found in the Appendix.
%~\ref{CTQMCapp}.
For the Kane-Mele-Hubbard model at half filling, there is no sign problem, even in the presence of $\pi$~fluxes.\cite{Hohenadler11}

We consider two observables to detect the Kondo screening of the spin fluxon:
the static spin susceptibility~$\chi_S$ and the local spectral function~$A_{\pi}(\omega)$.
To get access to $\chi_S$, we have extended the CT-INT method by implementing global susceptibility measurements.
Although $\hat{H}_U$ is confined to a small subset of the whole lattice,
we show in the Appendix
%~\ref{GlobSus}
how to measure $\chi_S$ on the whole lattice.
To get rid of the edges' contribution, we consider $\Delta\chi_S$ by subtracting the susceptibility of the same system without a $\pi$~flux and without interactions.
%In the following, we will always present $\Delta\chi_S$, where we subtract the susceptibility of the same system without $\pi$~flux
%and without interactions, to get rid of the edges' contribution.
%\begin{align}
% A(\nu,\omega) = \frac{1}{Z} \sum_{n,n'} \langle n | \hat{c}_{\nu} | n' \rangle   \langle n' | \hat{c}_{\nu}^{\dagger} | n \rangle 
%                 \left( e^{-\beta E_n} +  e^{-\beta E_{n'}} \right) \delta(\omega + E_n - E_{n'}),
%\end{align}
In contrast to $\chi_S$, the spectral function~$A_{\pi}(\omega) = -(1/6\pi) \sum_{i\in\hexagon} \operatorname{Im} \mathcal{G}_{i,i}^{R}(\omega)$
can be obtained from the onsite single-particle Green's function in imaginary time, $G_{i,i}(\tau)$, by averaging over the six sites of the flux-threaded plaquette
and afterwards performing the analytic continuation using the maximum entropy method.\cite{Beach2004,Abendschein06}
%Another important quantity to observe the Kondo effect is the single-particle spectral function~$A_{\pi}(\omega)$.
%It can be obtained from the imaginary-time single-particle Green's function.
%For the analytic continuation we use the maximum entropy method.\cite{Beach2004}

To observe the Kondo screening of the spin fluxon at low temperatures,
we have to pass the two characteristic energy scales of a Kondo system:
the valence fluctuation scale $k_BT_{\mathrm{VF}}$ and the Kondo scale $k_BT_K$.\cite{Coleman}
Starting from high temperatures, valence fluctuations are eliminated at $k_BT_{\mathrm{VF}}$
and a local moment is formed.
The impurity will stay in the local-moment regime until the temperature becomes lower than the Kondo temperature~$T_K$.
In the Kondo regime, the local moment is screened by the electronic bath via formation of a spin singlet.

\subsection{Local-moment formation\label{localmoment}}

\begin{figure}[b]
 \centering
 \begin{tabular}{cc}
 \hspace{0pt}\includegraphics[trim=0.2cm 0cm 0.1cm 0cm, clip=true, height=0.50\linewidth]{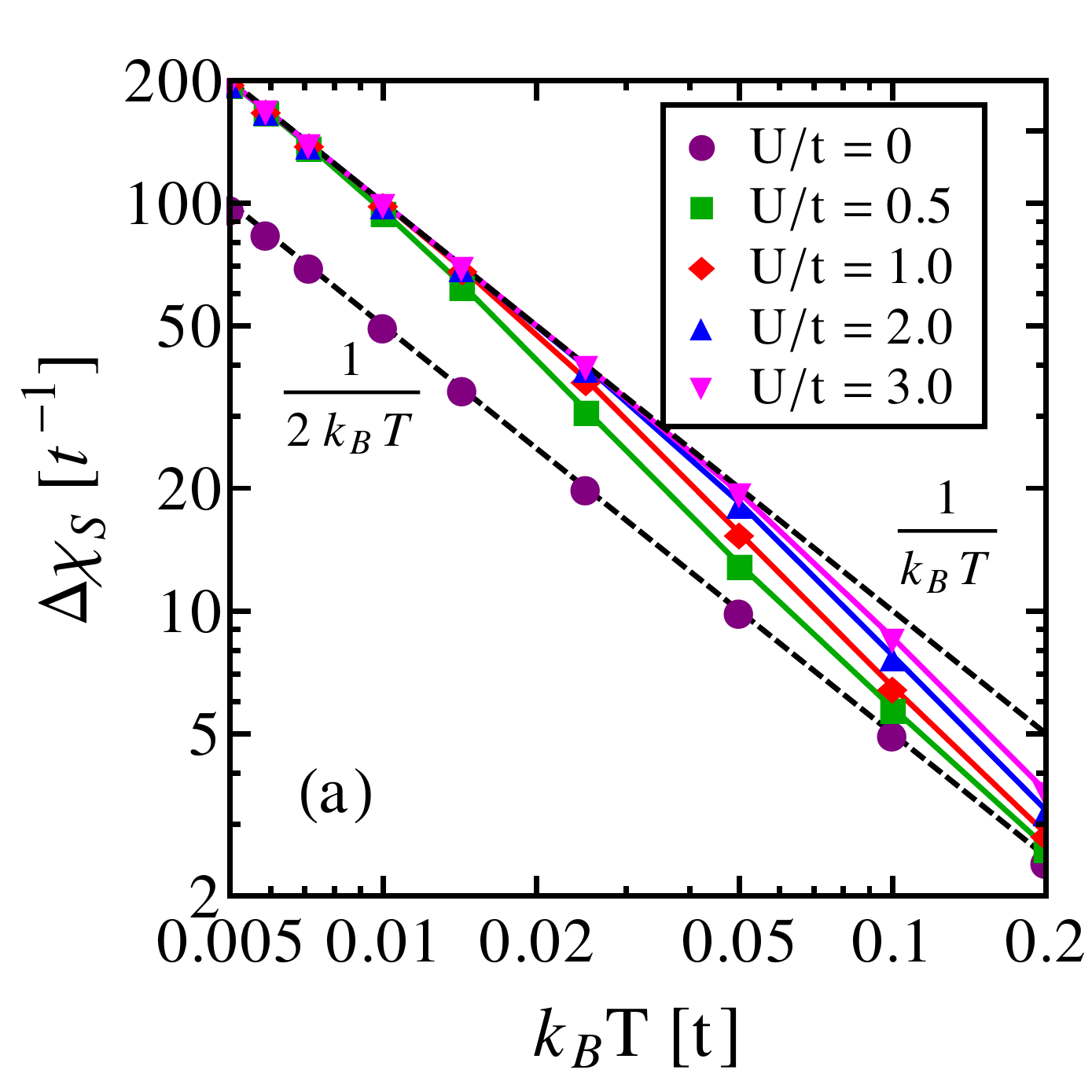} & \hspace{0pt}\includegraphics[trim=0.5cm 0cm 0.3cm 0cm, clip=true, height=0.5\linewidth]{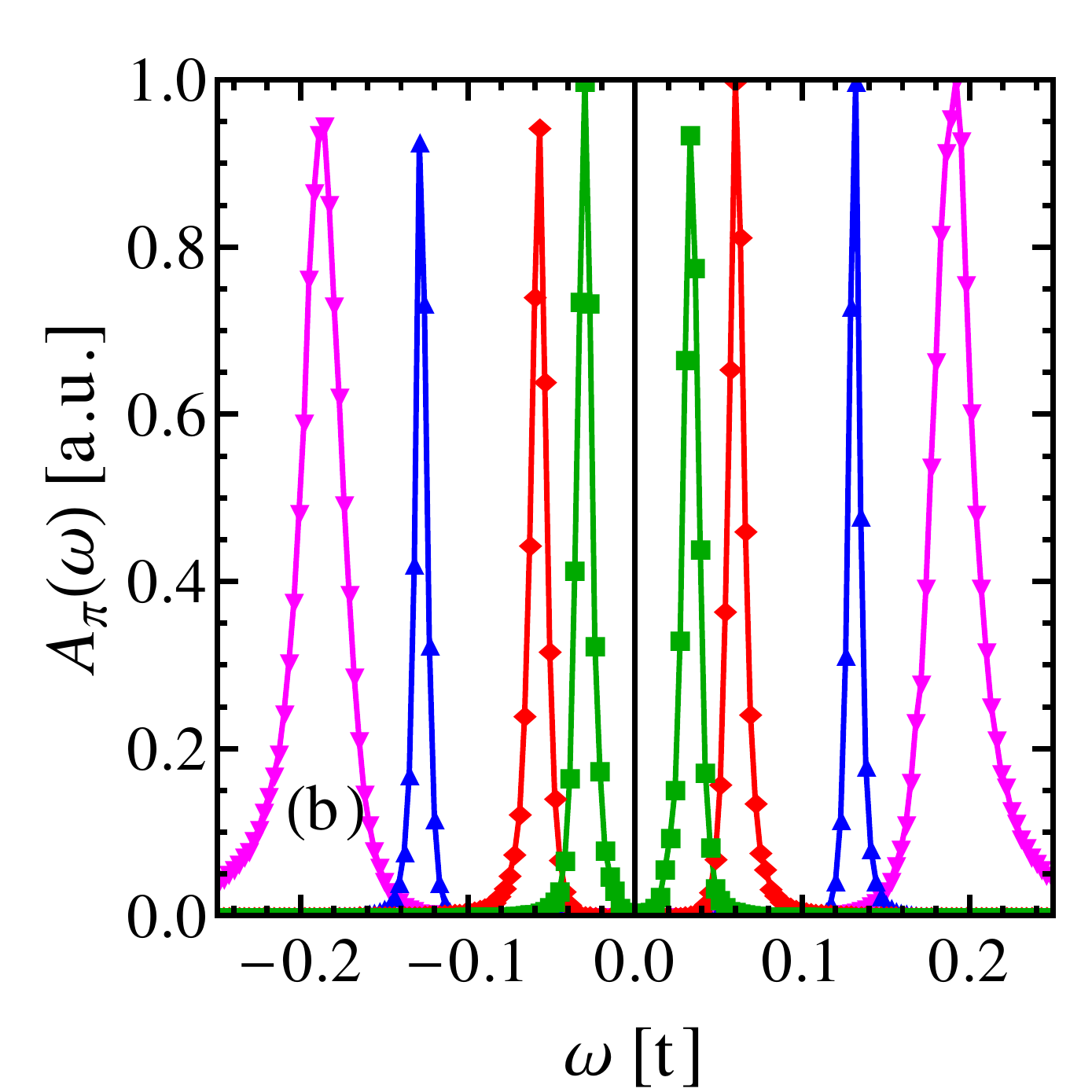}
 \end{tabular}
 \caption{(Color online) The spin susceptibility~$\Delta\chi_S$ (a) and the spectral function~$A_{\pi}(\omega)$ at $\beta t =100$ (b) of
          a $\pi$~flux at distance~$d=7$ from the edge of a ribbon ($W=15$, $L=500$, $\lambda/t=0.2$) are shown
					for several interaction strengths~$U$.
					$\Delta\chi_S$ is compared to the Curie law of a noninteracting ($\chi_S=1/2k_BT$)
					and an interacting system ($\chi_S=1/k_BT$).
          The error bars of $\Delta\chi_S$ are smaller than the symbol sizes and thus omitted.
          }
 \label{LocalMoments}
 \end{figure}

Before we can study the Kondo screening of the spin fluxon, we first have to determine the energy scale of local-moment formation.
To this end, we add $\hat{H}_U$ around a $\pi$~flux located deep in the bulk of a ribbon and slowly increase the interaction strength~$U$.
For $\lambda/t=0.2$, the results are presented in Fig.~\ref{LocalMoments}.
For all $U>0$, the spin susceptibility in Fig.~\ref{LocalMoments}(a) approaches the characteristic Curie law $\chi_S=1/k_BT$
of a $\pi$~flux in an interacting system at low temperatures.\cite{AssaadBercxHohenadler13}
This observation confirms that it is sufficient to include the Hubbard interaction only at the six sites of the flux-threaded plaquette.
However, with decreasing $U$, lower temperatures are necessary to converge to $\chi_S=1/k_BT$,
and for $U\rightarrow 0$ the susceptibility will stay at the $1/2k_BT$ Curie law of the noninteracting system.
In comparison to the single-impurity Anderson model,
where the valence fluctuation scale is given by $k_BT_{\mathrm{VF}}=U/2$,
here the local moment is formed at lower temperatures.
This fact can be related to the position of the Hubbard peaks in the spectral function~$A_{\pi}(\omega)$ [Fig.~\ref{LocalMoments}(b)].
The peak positions scale linearly with $U$, but the absolute energy scale is reduced by an order of magnitude.
In contrast to the single-impurity Anderson model, our impurity is not located at a single lattice site but is exponentially localized on the whole lattice.
As the largest amount of weight is equally distributed on the six lattice sites around the $\pi$~flux,
the valence fluctuation scale of the fluxon is approximately by a factor of $\gamma^2/6$ smaller.
Here, $\gamma$ is the fraction of the total weight that is located at the six sites of the flux-threaded hexagon in the noninteracting system.
For $\lambda/t=0.2$, we find $\gamma \approx 0.865$.

\subsection{Estimate of the Kondo temperature}

\begin{figure}[!htbc]
 \centering
 \begin{tabular}{cc}
 \hspace{-0pt}\includegraphics[height=0.54\linewidth]{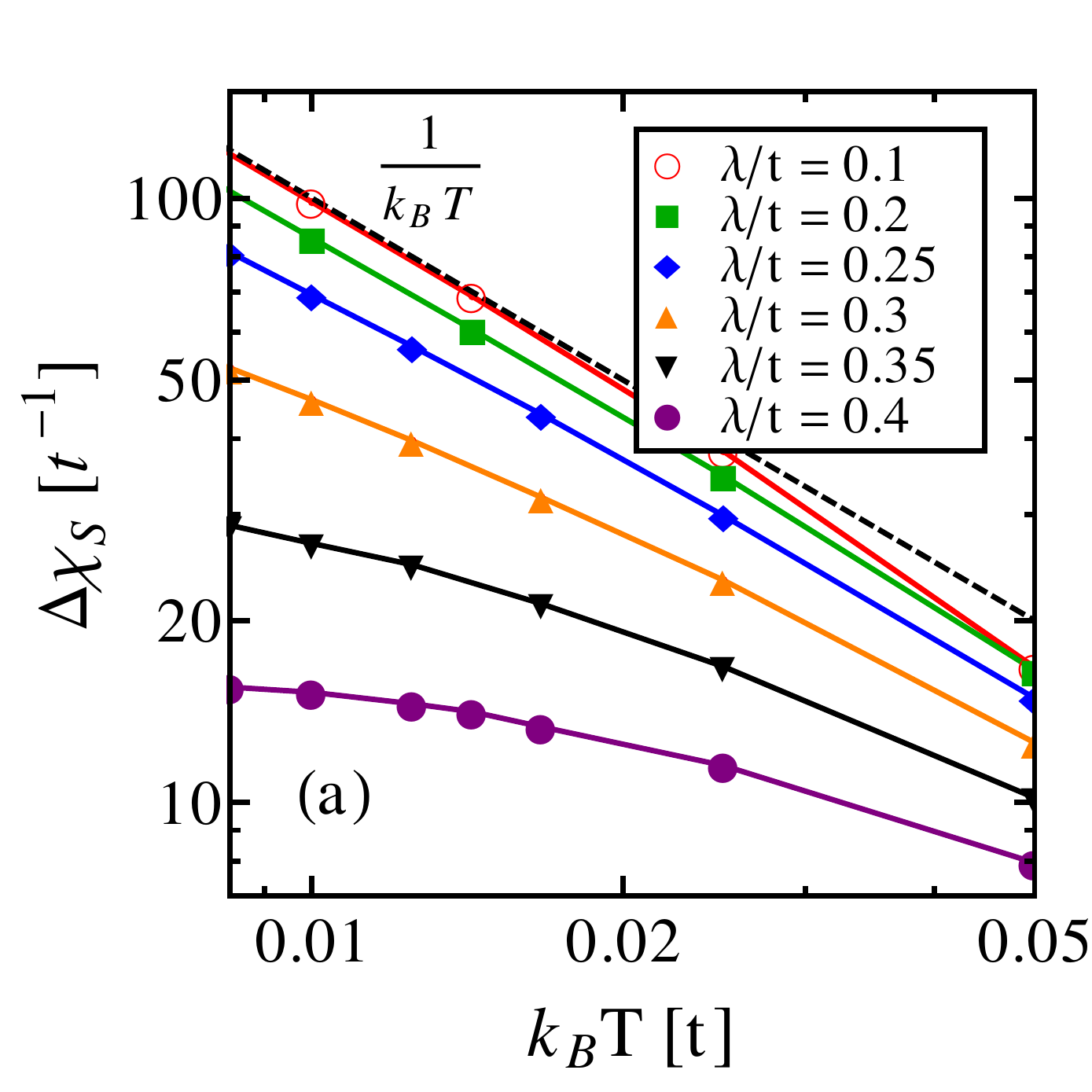} & \hspace{0pt}\includegraphics[trim=2.5cm 0cm 0cm 0cm, clip=true, height=0.54\linewidth]{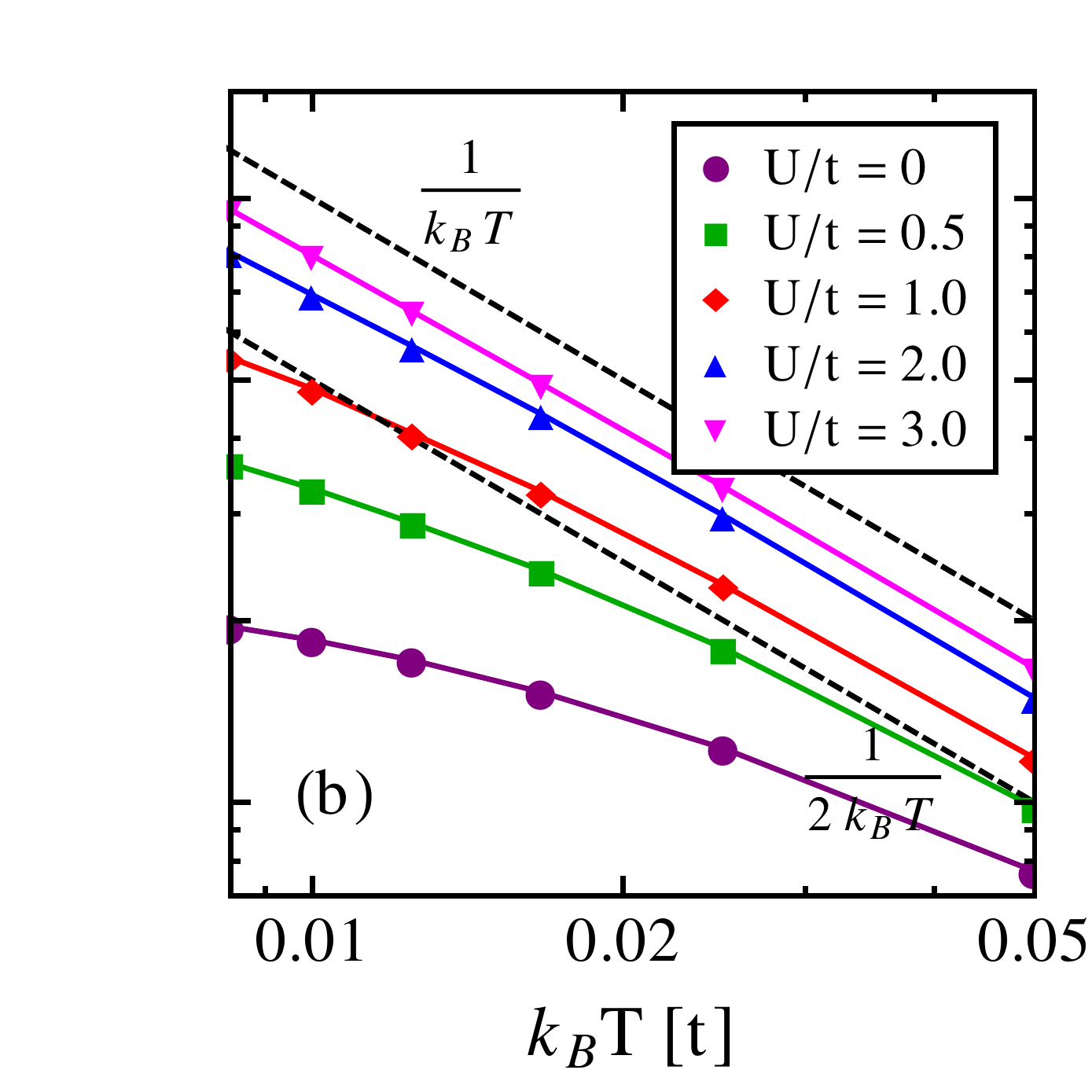}
 \end{tabular}
 \caption{(Color online) The spin susceptibility~$\Delta\chi_S$ of a $\pi$~flux at distance~$d=2$
          from the edge of a ribbon ($W=15$, $L=500$) shown in the atomic picture for $U/t=2$ (a)
					and in the adiabatic picture for $\lambda/t=0.25$ (b).
					For comparison, the Curie law of the (non)interacting system is drawn as a dashed line.
					%For an easier identification the data points of the Monte Carlo simulation are connected by a line.
					Error bars are smaller than the symbol sizes and thus omitted.
          }
 \label{CurieLawDev}
 \end{figure}

As the valence-fluctuation scale is rather low for our impurity model,
and the susceptibility measurements are affected by finite-size effects at low temperatures,
the energy window to detect the Kondo effect is only of the order of one magnitude.
Therefore, we have to maximize the hybridization by choosing a large spin-orbit coupling~$\lambda$
and the smallest possible distance $d=2$ between the $\pi$~flux and the edge.
The results for $\Delta\chi_S$ are shown in Fig.~\ref{CurieLawDev}.

There are two ways to present $\Delta\chi_S$: the atomic and the adiabatic picture.\cite{Coleman}
The atomic picture presented in Fig.~\ref{CurieLawDev}(a) considers a fixed $U/t=2$
and starts in the local-moment regime with a clear Curie-law behavior at $\lambda/t=0.1$.
With increasing hybridization~$\lambda$, the deviation from the Curie law becomes larger
as we enter the Kondo regime.
For $\lambda/t=0.4$, $\Delta\chi_S$ even seems to saturate at a constant value as $T\rightarrow0$.
In the adiabatic picture presented in Fig.~\ref{CurieLawDev}(b), one starts with the resonant state at $U/t=0$ for fixed $\lambda/t=0.25$.
It shows a large deviation from the noninteracting $1/2k_BT$ Curie law.
With increasing $U$, the susceptibility approaches the $1/k_BT$ Curie law as the system enters the local-moment regime.

\begin{figure}[t]
 \centering
 \hspace{0pt}\includegraphics[width=\linewidth]{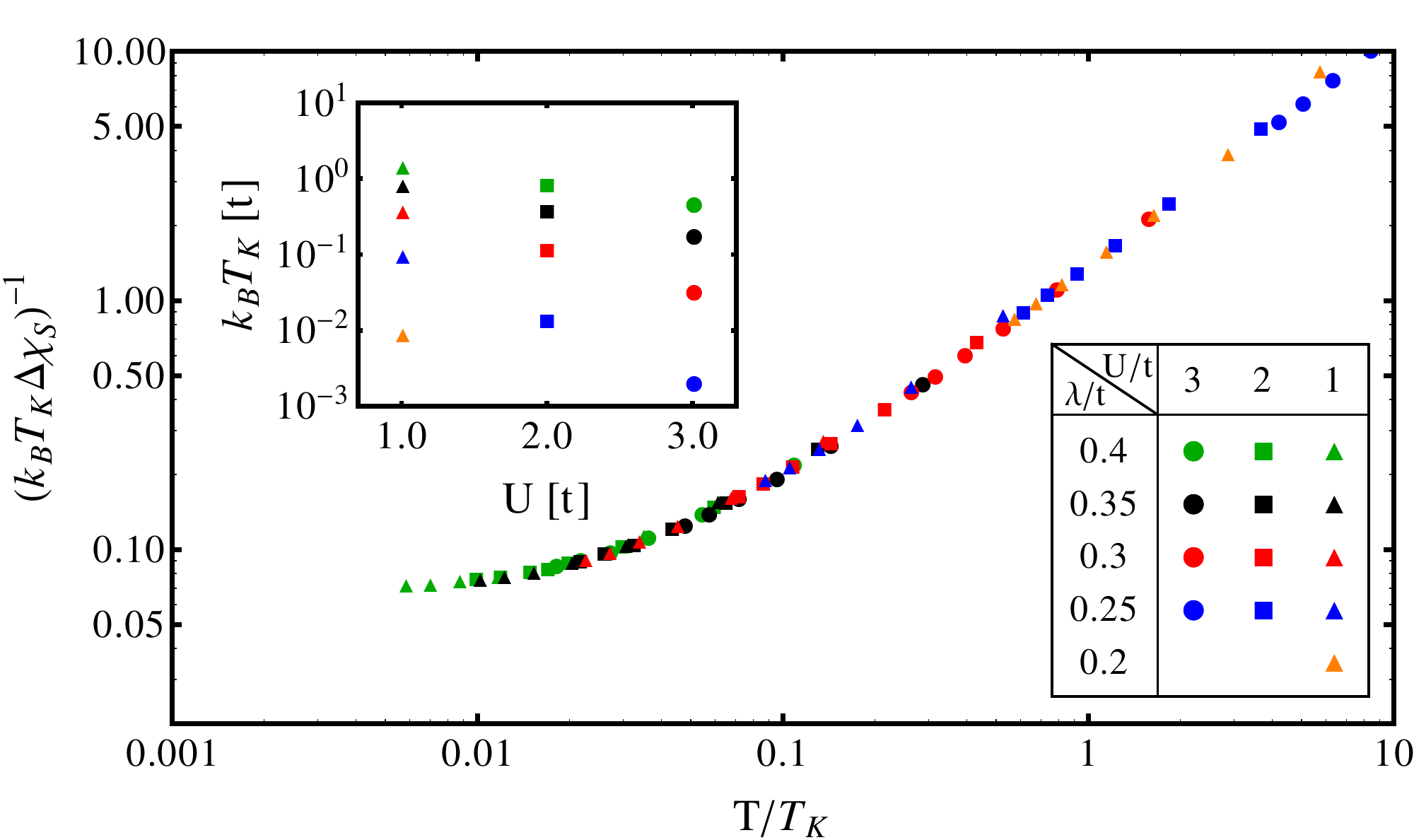}
 \caption{(Color online) Data collapse of the spin susceptibility $\Delta\chi_S$ for different values of~$U$ and~$\lambda$.
          The estimated Kondo temperatures~$T_K$ are shown in the inset.
          %Note that the data collapse only fixes the relative temperatures~$T_K$, but not their absolute values.
          We have taken the appearance of the Kondo resonance in Fig.~\ref{HubbHierarchy} as a reference to fix the absolute scale.
              }
 \label{DataCollapse}
 \end{figure}

At low temperatures, the Kondo temperature~$T_K$ becomes the only relevant energy scale,
leading to a scaling law $T_K \chi_S = \Phi(T/T_K)$ with a universal function~$\Phi(x)$.\cite{Coleman}
We use this scaling law to perform a data collapse of our susceptibility measurements, as shown in Fig.~\ref{DataCollapse}.
All the data points lie on a universal curve which deviates from the linear local-moment behavior at $T/T_K \ll 1$.
Since every data set is restricted to one order of magnitude in temperature, we use $13$ data sets to generate sufficient overlap.
Using the data collapse, we can estimate the Kondo temperature of our system which is plotted in the inset of Fig.~\ref{DataCollapse}.

\subsection{Kondo resonance}

The hierarchy of energy scales can be best captured by looking at the local spectral function~$A_{\pi}(\omega)$ of the fluxon as a function of temperature.
 \begin{figure}[!b]
 \centering
 \begin{tabular}{ccc}
 \hspace{0pt}\includegraphics[height=0.55\linewidth]{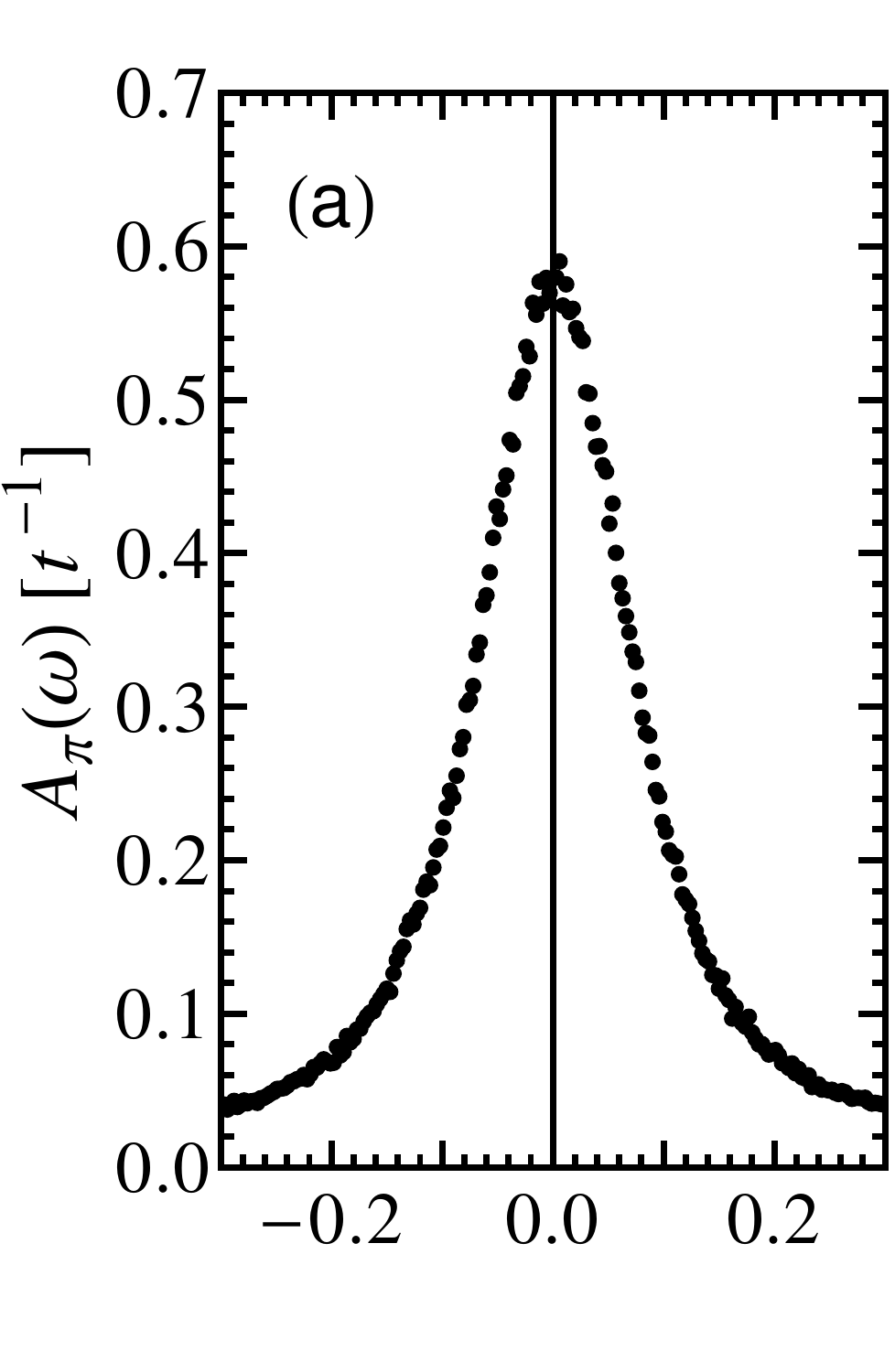} & \hspace{-0pt}\includegraphics[trim=2.2cm 0cm 0cm 0cm, clip=true, height=0.55\linewidth]{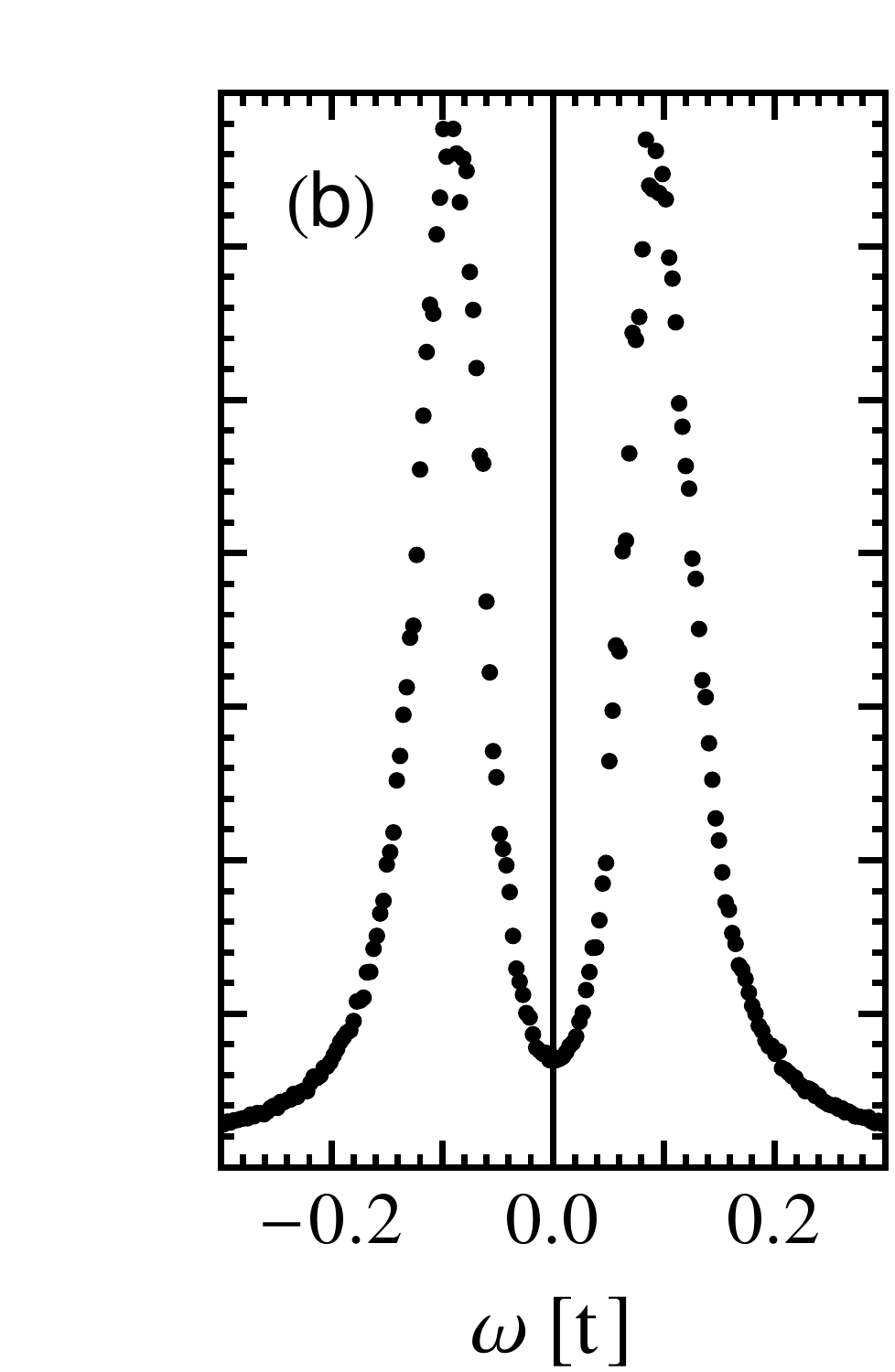} &
 \hspace{-0pt}\includegraphics[trim=2.2cm 0cm 0cm 0cm, clip=true, height=0.55\linewidth]{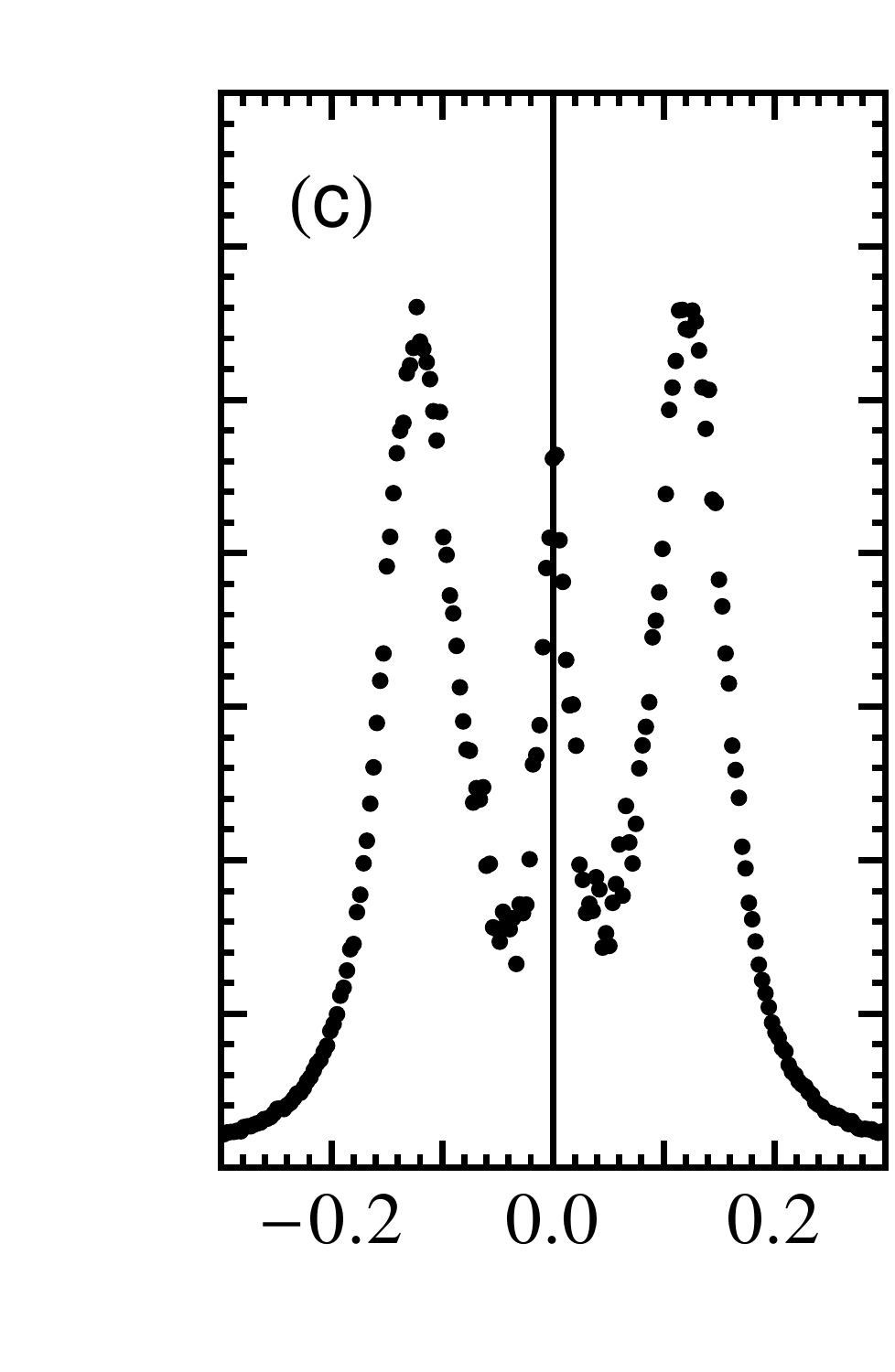}
 \end{tabular}
 \caption{The local spectral function~$A_{\pi}(\omega)$ of a $\pi$~flux at distance~$d=2$ to the edge of the ribbon ($W=15$, $L=500$, $\lambda/t=0.25$, $U/t=2$) is shown for temperatures $\beta t = 10$ (a), $\beta t = 40$ (b), and $\beta t=100$ (c).
              }
 \label{HubbHierarchy}
 \end{figure}
$A_{\pi}(\omega)$ is shown in Fig.~\ref{HubbHierarchy} at fixed $U/t=2$ and $\lambda/t=0.25$ for the three different regimes of the Kondo problem.
At $\beta t=10$, the impurity is in the high-temperature regime where all correlation effects are washed out by thermal fluctuations.
Thus, there is only a single broad peak at $\omega=0$ visible in Fig.~\ref{HubbHierarchy}(a).
As the impurity enters the local-moment regime shown in Fig.~\ref{HubbHierarchy}(b) for $\beta t =40$,
the single peak splits symmetrically into the two Hubbard peaks already observed in Fig.~\ref{LocalMoments}(b).
The peak positions are given by the energy necessary to excite the spin fluxon to a charge fluxon.
Further lowering the temperature leads to the Kondo regime, as shown in Fig.~\ref{HubbHierarchy}(c) for $\beta t =100$,
where the Kondo resonance emerges at $\omega=0$.

\section{Conclusions\label{conclusions}}

%In this paper, we have considered the fluxon states created by a $\pi$~flux as a Kondo impurity that is screened by the helical edge liquid of a QSHI.

%In this paper, we have used the fluxon states due to a $\pi$~flux inserted into a QSHI to construct an impurity model

%In this paper, we have used the fluxon states created by a $\pi$~flux to construct a Kondo impurity model
%where the spin fluxon is screened by the helical edge liquid of a QSHI.

In this paper, we have considered a realization of the Kondo effect in a QSHI, where the spin fluxon states
created by a $\pi$~flux are screened by the helical edge states.
Compared to ordinary impurity models, the spin fluxon states play the role of the free spin,
whereas the helical edge liquid serves as the conduction band.
As the fluxon is an extended object emerging inside the lattice of the QSHI,
the analysis of the Kondo effect requires a good understanding of both hybridization and local-moment formation.

%In this paper, we have constructed a Kondo impurity model where the spin fluxon states
%created by a $\pi$~flux are screened by the helical edge states of the Kane-Mele-Hubbard model.
%The study of this model was complicated by the fact that the fluxon is an extended object
%emerging inside the lattice of the QSHI.
%Thus, the analysis of the Kondo effect requires a good understanding of both hybridization and local-moment formation.

Since there is no explicit hybridization parameter in our model, we have first studied the hybridization between fluxon and edge states
of the Kane-Mele model with zigzag terminations depending on their distance~$d$ and the spin-orbit couling~$\lambda$.
We have found that there is only a finite hybridization at even~$d$,
where it monotonically increases with increasing $\lambda$.
A comparison with the extent of both fluxon and edge states showed that their hybridization is dominated by the extent of the edge states.
The odd-even effect with $d$ is connected to the coupling between opposite edge states, where it also appears as a function of distance.

To study the impurity model in the presence of repulsive Hubbard interactions, we have used the CT-INT method,
extended by global susceptibility measurements.
We have shown that it is sufficient to include the interaction terms only directly around the $\pi$~flux
in order to obtain the characteristic Curie law of the spin fluxons at low temperatures.
The local spectral function at the flux-threaded plaquette exhibits two Hubbard peaks,
which correspond to excitations between spin and charge fluxons and define the energy scale of the local-moment formation.
However, due to the finite extent of the fluxons, the peak positions appear at much lower energies than in the single-impurity Anderson model.

We have studied the Kondo screening of the spin fluxon at distance $d=2$ from the edge to maximize the hybridization.
When entering the Kondo regime, the spin susceptibility deviates from the Curie law following the universal behavior obtained from a data collapse.
Thereby, we have estimated the Kondo temperature for different values of $\lambda$ and $U$.
As a function of temperature, the local spectral function shows the characteristic signatures of the high-temperature, local-moment, and Kondo regimes,
with the Kondo resonance emerging at low temperatures.

Here, we have studied the Kondo screening of the spin fluxon for a $\pi$~flux confined to a single plaquette.
However, fluxon states are also present for an extended magnetic $\pi$~flux.\cite{QiZhang08}
As soon as the $\pi$~flux extends beyond a single plaquette, time-reversal symmetry is locally broken
and the midgap states of the noninteracting system split depending on the spin sector.\cite{QiZhang08}
Thus, also the two spin fluxon states lose their degeneracy.
As long as the corresponding energy splitting is smaller than the Kondo scale, we still expect to see the Kondo effect with a single Kondo resonance,
but with increasing extent of the $\pi$~flux the Kondo resonance will split into two parts.
Ultimately, the Kondo effect will disappear when the extent of the magnetic flux is too large.
Finally, the $\pi$~flux is not the only defect that could show the Kondo effect:
Lattice dislocations also lead to localized states inside the bulk band gap.\cite{Juricic12}

\begin{acknowledgments}

We are grateful to F.~Goth, T.~Müller, and M.~Sprengel for helpful discussions
and thank the Jülich Supercomputing center for generous allocation of CPU time.
We acknowledge financial support from the DFG Grant Nos. AS120/10-1 and Ho~4489/2-1 (FOR 1807).

\end{acknowledgments}

\appendix*

%\section{Extended $\pi$~flux\label{extendedPi}}

\section{Global susceptibility measurements within the CT-INT method\label{CTQMCapp}}

%Impurity problems, as discussed in this paper, only contain a few interacting degrees of freedom that are coupled to a large noninteracting bath.
%In an action-based formulation all the bath degrees of freedom can be integrated out, which results in an effective action only defined on the interacting lattice sites.
%The CT-INT method introduced in Ref.~\onlinecite{Rubtsov05} makes use of this effective action to efficiently simulate the physical problem.
%%However, it only allows to calculate observables locally on the remaining degrees of freedom.
%However, the calculation of observables will only be practicable locally on the remaining degrees of freedom.
%In the following, we will show how to calculate global observables that are defined on the whole lattice.

Within
%the action-based formulation of
the CT-INT method, all the bath degrees of freedom can be integrated out,
resulting in an effective action only defined on the interacting lattice sites.
In the following, we will show how to calculate global observables that are defined on the whole lattice.

\subsection{Hamiltonian}

Consider the many-body Hamiltonian
\begin{align}
\hat{H} = \hat{H}_0 + \hat{H}_U,
\label{totalHamilton}
\end{align}
where $\hat{H}_0$ is the free and~$\hat{H}_U$ the interacting part.
We assume that $\hat{H}_0$ conserves both total particle number and spin, \ie,
%We will look at
%that consists of a noninteracting part
\begin{align}
\hat{H}_0 = \sum_{i,j\in\Omega} \sum_{\sigma} \hat{c}^{\dagger}_{i,\sigma} \mathcal{H}^{\sigma}_{i,j} \hat{c}^{\phantom{}}_{j,\sigma},
\label{freeHamilton}
\end{align}
where $\Omega$ is the set of all lattice sites.
$\hat{H}_0$ can be diagonalized by a unitary matrix~$U$ that is block diagonal in spin space and transforms the annihilation operator as
$\hat{c}_{j,\sigma} = \sum_{m} U_{j,m,\sigma} \hat{\gamma}_{m,\sigma}$.
The Hamiltonian~(\ref{freeHamilton}) becomes
$\hat{H}_0 = \sum_{m,\sigma} \epsilon_{m,\sigma} \hat{\gamma}^{\dagger}_{m,\sigma} \hat{\gamma}^{\phantom{}}_{m,\sigma}$
with eigenvalues $\epsilon_{m,\sigma}$.
Using this notation, the noninteracting single-particle Green's function can be written as
\begin{align}
\begin{split}
G^{0,\sigma}_{i,j}(\tau) &= \langle T \hat{c}_{i,\sigma}^{\dagger}(\tau) \hat{c}_{j,\sigma} \rangle_0 \\
&= \sum\limits_{m} U_{j,m,\sigma}^{\phantom{}} \, U_{m,i,\sigma}^{\dagger} \, \frac{e^{\tau\epsilon_{m,\sigma} }}{1 + e^{\beta \epsilon_{m,\sigma}}}
\label{greens}
\end{split}
\end{align}
for $0 \leq \tau < \beta$.
Here, $\langle \bullet \rangle_0 = \Tr [e^{-\beta \hat{H_0} } \bullet ]/Z_0$ is the expectation value with respect to $\hat{H}_0$,
and $T$ is the time-ordering operator.

In the following calculation, $\hat{H}_U$ can be a general interaction term, where the lattice sites are restricted to a subset~$\mathcal{S}\subseteq\Omega$.
We consider a repulsive Hubbard interaction of the form
\begin{align}
\hat{H}_U = \frac{U}{2} \sum_{i\in\mathcal{S}} \sum_{s=\pm1} \left[ \hat{n}_{i,\uparrow} - \alpha_{\uparrow}(s) \right] \left[ \hat{n}_{i,\downarrow} - \alpha_{\downarrow}(s) \right]
%\hat{H}_U = \sum_{i,j,k,l\in\mathcal{S}} \sum_{\sigma,\sigma',\sigma'',\sigma'''} \hat{c}^{\dagger}_{i,\sigma} \hat{c}^{\dagger}_{j,\sigma'} \hat{c}_{k,\sigma''} \hat{c}_{l,\sigma'''}.
\label{HubbardAlpha}
\end{align}
with $\alpha_{\sigma}(s) = 1/2 + \sigma s \delta$ and $\delta = 1/2 + 0^{+}$.
The Ising spin variable~$s$ allows us to avoid the negative-sign problem.\cite{AssaadLang07}

\subsection{Calculation of observables}

In this section, we give a short outline of how to calculate observables in the CT-INT method.
We will use the notation of Ref.~\onlinecite{AssaadLang07}; for a review see Ref.~\onlinecite{GullReview}.

Monte Carlo methods rely on the possibility of rewriting the expectation value of an observable~$\hat{O}$ as
\begin{align}
\langle\hat{O}\rangle = \sum_{C} W(C) \llangle \hat{O} \rrangle_{C},
\label{MCobservable}
\end{align}
where the sum runs over all possible configurations~$C$.
Here, $W(C)$ has to be a probability distribution, and $\llangle \bullet \rrangle_C$ denotes the expectation value for a single configuration.
In the Monte Carlo process, the sum is sampled by picking configurations according to $W(C)$, so that only $\llangle \hat{O} \rrangle_{C}$ has to be evaluated.

The CT-INT method is based on a weak-coupling perturbation expansion of $\hat{H}_U$ with respect to the free Hamiltonian~$\hat{H}_0$.
The notation needed below can be most easily explained by writing down the partition function for the Hubbard interaction~(\ref{HubbardAlpha}),
\begin{align}
 \frac{Z}{Z_0} = \sum_{C_n} \left(- \frac{U}{2} \right)^n \det M_{\sigma}(C_n).
\label{PartitionFunction}
\end{align}
The sum over all configurations is defined as
\begin{align}
\sum_{C_n} = \sum_{n=0}^{\infty} \int\limits_0^{\beta} d\tau_1 \sum_{i_1,s_1} \cdots \int\limits_0^{\tau_{n-1}} d\tau_n \sum_{i_n,s_n}.
\label{sum}
\end{align}
A single configuration $C_n = \{[i_1,\tau_1,s_1], \dots, [i_n,\tau_n,s_n]\}$ is specified by the perturbation order~$n$ defining $n$ vertices,
each containing a lattice site, imaginary time, and Ising spin variable.
The matrix $M_{\sigma}(C_n)$ is defined via Wick's theorem, which holds for expectation values $\langle \bullet \rangle_0$ and leads to the relation
\begin{align}
\det M_{\sigma}(C_n) = \langle T \prod_{k=1}^{n} \left[\hat{n}_{i_k,\sigma}(\tau_k) -\alpha_{\sigma}(s_k) \right] \rangle_0.
\end{align}
%The basic notation introduced here stays true for a more general interaction term, but the vertex variables and weights may change.

%Within the CT-INT method, observables are calculated using Eq.~(\ref{MCobservable}).
%The weight $W(C_n)$ used for the sampling can be found in Ref.~\onlinecite{AssaadLang07}.
Wick's theorem also holds for $\llangle \hat{O} \rrangle_{C_n}$, since only expectation values with respect to $\hat{H}_0$ appear in the perturbation expansion.\cite{Luitz10}
Thus, every expectation value can be calculated from the single-particle Green's function
\begin{align}
\begin{split}
\llangle T \hat{c}^{\dagger}_{i,\sigma}(&\tau) \hat{c}_{j,\sigma}(\tau') \rrangle_{C_n}
= G^{0,\sigma}_{i,j}(\tau,\tau') \\
&- \sum_{r,s=1}^n G^{0,\sigma}_{i,i_r}(\tau,\tau_{r})(M_{\sigma}^{-1})_{r,s} G^{0,\sigma}_{i_s,j}(\tau_{s},\tau')
\end{split}
\label{GF_Cn}
\end{align}
for a configuration~$C_n$.

\subsection{Global susceptibility measurements\label{GlobSus}}

Observables like the spin susceptibility can be calculated for each configuration~$C_n$ using Wick's theorem and the Green's function~(\ref{GF_Cn}).
For an interaction term $\hat{H}_U$ only defined locally on a subset~$\mathcal{S}$ of the lattice, one would measure the local spin susceptibility
\begin{align}
\chi_S^l = \int\limits_0^{\beta} \langle \hat{S}_z^l(\tau) \hat{S}_z^l \rangle \, d\tau 
\end{align}
with the local spin operator $\hat{S}_z^l = \sum_{i\in\mathcal{S}} \sum_{\sigma} \sigma \hat{c}^{\dagger}_{i,\sigma} \hat{c}_{i,\sigma}$.
However, Eq.~(\ref{GF_Cn}) is valid for all lattice sites in $\Omega$.
Therefore, it is possible to calculate the static spin susceptibility (\ref{spinsus})
%\begin{align}
%\chi_S = \beta \left( \langle \hat{S}_z^2 \rangle - \langle \hat{S}_z \rangle^2 \right)
%\label{chi_app}
%\end{align}
for a spin operator~$\hat{S}_z$ defined on the whole lattice.
In principle, one can first calculate the Green's function~(\ref{GF_Cn}) between two lattice sites and afterwards $\chi_S$,
but for large system sizes this becomes unfeasible.
In the following, we will show how $\chi_S$ can be computed by first performing the sum over all lattice sites.
%This will define new objects that will simplify the calculation of the susceptibility.

As our Hamiltonian~(\ref{totalHamilton}) has a $U(1)$~spin symmetry, $\langle\hat{S}_z\rangle = 0$ and we only need to calculate $\langle\hat{S}_z^2\rangle$.
The corresponding expectation value for Eq.~(\ref{spinsus}) is given by
\begin{align}
\begin{split}
\llangle \hat{S}_z^2 \rrangle_{C_n} = %\llangle \hat{S}_z \rrangle_{C_n}^2 \\
\sum_{i,j} \sum_{\sigma,\sigma'} \sigma \sigma' \llangle T \hat{c}_{i,\sigma}^{\dagger} \hat{c}_{i,\sigma}  \rrangle_{C_n}
\llangle T \hat{c}_{j,\sigma'}^{\dagger} \hat{c}_{j,\sigma'}  \rrangle_{C_n} \\
%[ \sum_{i\in\Omega} \sum_{\sigma} \sigma \llangle T \hat{c}_{i,\sigma}^{\dagger} \hat{c}_{i,\sigma}  \rrangle_{C_n} ]^2 \\
+ \sum_{i,j} \sum_{\sigma} \llangle T \hat{c}_{i,\sigma}^{\dagger} \hat{c}_{j,\sigma}  \rrangle_{C_n} \left( \delta_{i,j}
- \llangle T \hat{c}_{j,\sigma}^{\dagger} \hat{c}_{i,\sigma}  \rrangle_{C_n}    \right),
\end{split}
\label{wickd}
\end{align}
where we have already used Wick's theorem.
Note that $i,j\in\Omega$.
The first term is just $\llangle \hat{S}_z \rrangle_{C_n}^2$ and can be rewritten by inserting Eq.~(\ref{GF_Cn}) to arrive at
\begin{align}
\llangle \hat{S}_z \rrangle_{C_n}
= \langle \hat{S}_z \rangle_0
- \sum_{\sigma} \sigma \Tr\left[\Gamma_{\sigma} M_{\sigma}^{-1}\right].
\label{spincor}
\end{align}
The trace is over all $n$~vertices of the configuration~$C_n$, and the $n\times n$~matrix $\Gamma_{\sigma}$ is given by
\begin{align}
\begin{split}
\Gamma_{\sigma,s,r}
&= \Gamma_{\sigma,i_s,i_r}(\tau_s - \tau_r) 
= \sum_{i\in\Omega} G^{0,\sigma}_{i_s,i}(\tau_s) \, G^{0,\sigma}_{i,i_r}(-\tau_r) \\
%&= \sum_{i\in\Omega} \langle T \hat{c}_{i_s,\sigma}^{\dagger}(\tau_{s}) \hat{c}_{i,\sigma} \rangle_0 \langle T \hat{c}_{i,\sigma}^{\dagger} \hat{c}_{i_r,\sigma}(\tau_{r}) \rangle_0 \\
&= - \frac{1}{2} \sum_m U_{i_r,m,\sigma}^{\phantom{}} \, U_{m,i_s,\sigma}^{\dagger} \, \frac{e^{(\tau_{s}-\tau_{r})\epsilon_{m,\sigma}}}{1+\cosh(\beta \epsilon_{m,\sigma})}.
\end{split}
\label{gamma}
\end{align}
For the calculation, we have to insert the noninteracting Green's function~(\ref{greens}) using its antiperiodicity for negative time arguments
%The case $\tau_r=0$ does not cause any problems since the probability of choosing exactly $\tau_r=0$ is zero.
and the unitarity of $U_{\sigma}$ to get rid of the sum over all lattice sites $i\in\Omega$.

A similar calculation for the second part of Eq.~(\ref{wickd}) requires the introduction of another $n\times n$ matrix,
\begin{align}
\begin{split}
\Delta_{\sigma,s,r} &= \Delta_{\sigma,i_s,i_r}(\tau_s - \tau_r) 
%&= \sum_{i,j\in\Omega} \langle T \hat{c}_{i_s,\sigma}^{\dagger} (\tau_{s}) \hat{c}_{i,\sigma} \rangle_0
%\langle \hat{c}_{i,\sigma}^{\dagger} \hat{c}_{j,\sigma} \rangle_0
%\langle T \hat{c}_{j,\sigma}^{\dagger} \hat{c}_{i_r,\sigma} (\tau_{r}) \rangle_0 \\
= \sum_{i,j\in\Omega} G^{0,\sigma}_{i_s,i}(\tau_s) \, G^{0,\sigma}_{i,j}(0) \, G^{0,\sigma}_{j,i_r}(-\tau_r) \\
&= - \frac{1}{2} \sum_m
\frac{U_{i_r,m,\sigma}^{\phantom{}} \, U_{m,i_s,\sigma}^{\dagger} \, e^{(\tau_{s}-\tau_{r})\epsilon_{m,\sigma}}}{\left[1 + \cosh(\beta \epsilon_{m,\sigma})\right]\left( 1+e^{\beta \epsilon_{m,\sigma}}\right)}.
\end{split}
\end{align}

All in all, we have
\begin{align}
\begin{split}
\llangle \hat{S}_z^2 \rrangle_{C_n} = & \langle \hat{S}_z^2 \rangle_0
-  \langle \hat{S}_z\rangle_0^2
+ \sum_{\sigma} \Tr \left[ (2 \Delta_{\sigma} - \Gamma_{\sigma})  M_{\sigma}^{-1} \right] \\
& + \left( \langle \hat{S}_z\rangle_0 -  \sum_{\sigma} \sigma \Tr\left[\Gamma_{\sigma} M_{\sigma}^{-1}\right]  \right)^2 \\
&- \sum_{\sigma} \Tr \left[ \Gamma_{\sigma} M_{\sigma}^{-1} \Gamma_{\sigma} M_{\sigma}^{-1} \right].
\end{split}
\label{tracey}
\end{align}
The computational effort for calculating $\chi_S$ in this way is dominated by the last term in Eq.~(\ref{tracey}).
The evaluation of the matrix product $\Gamma_{\sigma} M_{\sigma}^{-1}$ is $\mathcal{O}(n^3)$.
It is typically more expensive than local, time-dependent measurements, but it still scales like the original CT-INT method.

Our calculation for $\chi_S$ can easily be transferred to the
charge susceptibility $\chi_C = \beta ( \langle \hat{N}^2 \rangle - \langle \hat{N} \rangle^2 )$,
where $\hat{N}$ is the total particle number operator.
One only has to substitute $\hat{S}_z\rightarrow\hat{N}$ and $\sigma \rightarrow 1$ in Eq.~(\ref{tracey}).

%\bibliography{database}

%Merlin.mbs v4.21 2009-07-09.
%

\end{document}